\documentstyle[psfig]{mn2e}

\begin{document}

\title[SCUBA Observations of Dust around Lindroos Stars]
  {SCUBA Observations of Dust around Lindroos Stars:
   Evidence for a Substantial Submillimetre Disc Population}

\author[M. C. Wyatt et al.]
  {M.~C.~Wyatt\thanks{Email: wyatt@roe.ac.uk},
   W.~R.~F.~Dent and J.~S.~Greaves\\
  UK Astronomy Technology Centre, Royal Observatory,
  Blackford Hill, Edinburgh EH9 3HJ}

\maketitle

\begin{abstract}
We have observed 22 young stars from the Lindroos sample (Lindroos 1986)
at 850 $\mu$m with SCUBA on the JCMT to search for evidence of dust discs.
Stars in this sample are the less massive companions of B-type primaries
and have well defined ages that are $10-170$ Myr;
i.e., they are about to, or have recently arrived on the main
sequence.
Dust was detected around three of these stars (HD74067, HD112412 and 
HD99803B).
The emission around HD74067 is centrally peaked and is approximately
symmetrically distributed out to $\sim 70$ arcsec from the star.
This emission either arises from a two component disc, one 
circumstellar and the other circumbinary with dust masses of 0.3
and $>27M_\oplus$ respectively, or an unrelated background object.
The other two detections we attribute to circumsecondary discs
with masses of 0.04 and 0.3 $M_\oplus$;
we were also able to show that a circumprimary disc is present
around HD112413 with a similar mass to that around the companion
HD112412.
Cross-correlation of our sample with the \textit{IRAS} catalogs only
showed evidence for dust emission at 25 $\mu$m and 60 $\mu$m toward
one star (HD1438);
none of the sub-mm detections were evident in the far-IR data implying
that these discs are cold ($< 40$ K assuming $\beta=1$).
Our sub-mm detections are some of the first of dust discs surrounding
evolved stars that were not detected by \textit{IRAS} or \textit{ISO}
and imply that 9-14\% of stars could harbour previously undetected
dust discs that await discovery in unbiased sub-mm surveys.
If these discs are protoplanetary remnants, rather than secondary
debris discs, dust lifetime arguments show that they must be devoid
of small $<0.1$ mm grains.
Thus it may be possible to determine the origin of these discs from
their spectral energy distributions once these have been better
defined.
The low inferred dust masses for this sample support the picture that 
protoplanetary dust discs are depleted to the levels of the brightest 
debris discs ($\sim 1 M_\oplus$) within 10 Myr, although if the
extended emission of HD74067 is associated with the star, this would
indicate that $>10M_\oplus$ of circumbinary material can persist
until $\sim 60$ Myr and would also support the theory that T Tauri
discs in binary systems are replenished by circumbinary envelopes.
\end{abstract}

\begin{keywords}
  circumstellar matter --
  stars: binaries: visual --
  stars: planetary systems: protoplanetary discs --
  stars: pre-main-sequence --
  submillimetre.
\end{keywords}

\section{Introduction}
\label{sec-intro}
Most, if not all, stars are born with a disc
of gas and dust (e.g., Shu, Adams \& Lizano 1987).
The fate of such protoplanetary discs is important, since it is
within these discs that planets are thought to form through
the coagulation of initially sub-$\micron$-sized
dust grains (Weidenschilling \& Cuzzi 1993; Lissauer 1993).
Such grain growth is expected to continue as long as the parent disc
remains sufficiently massive and models show that our solar system's
terrestrial planets could have achieved a sizeable fraction of their
final mass by this mechanism within 10 Myr (e.g., Lissauer 1993).
However, at the same time as grain growth is occurring several other
mechanisms are competing to deplete disc material.
These mechanisms include:
viscous gas drag and Poynting-Robertson light drag, both of which
result in the accretion of disc material onto the star;
stripping of disc material by a stellar wind and radiation pressure;
removal of material during encounters with nearby stars;
and photoevaporation, either by the parent star or by an external star
(Clarke, Gendrin, \& Sotomayor 2001; Hollenbach, Yorke \&
Johnstone 2000).
Clearly the success or otherwise of planet formation depends
on whether this process is complete at the time the disc becomes
too depleted for planetesimal growth.
In this regard an important observable is how
long stars retain their protoplanetary discs.

Observations of the near-IR excess emission from stars in nearby
open clusters show that the fraction of stars with detectable discs
decreases rapidly with the age of the cluster from around $>80$\% for
recently formed clusters, down to 0\% for 6 Myr-old clusters (Haisch,
Lada, \& Lada 2001).
Similarly no near-IR excess is found in nearby star forming regions
for stars older than 3-10 Myr (Strom et al. 1989; Kenyon \& Hartmann
1995).
However, a near-IR excess only implies the presence of warm dust that
resides closer to the star than a few 0.1 AU and it is still possible
that massive outer discs exist in the older systems without material
in the central AU.
The cool outer portions of the discs are best probed using
sub-mm/mm observations.
Such observations are particularly useful, because they also result in a
reliable estimate of the disc mass.
This is because the discs are transparent at these wavelengths (so
all the mass is observed), and the mass estimates are relatively unaffected by
uncertainties in the contribution of the stellar photosphere, as well as
being only weakly dependent on estimates of the size of the dust grains
and of their emitting temperature (e.g., Zuckerman 2001; see section 6).

However, so far sub-mm/mm studies of the evolution of disc mass in
the first 10 Myr have been inconclusive.
Such observations show that while the masses of the dust discs
(where present) are similar for stars of all spectral types at 30-70
$M_\oplus$ (Mannings \& Sargent 1997), it is not clear whether this disc
mass does (N\"{u}rnberger et al. 1997) or does not (Osterloh \&
Beckwith 1995) decrease with age.
The evidence is also inconclusive as to whether the decay of the sub-mm
disc is (Skinner, Brown, \& Walter 1991; Andre \& Montmerle 1994;
Duvert et al. 2000) or is not (Beckwith et al. 1990; Osterloh \& Beckwith
1995; N\"{u}rnberger et al. 1998) coincident with the disappearance of the
inner disc.
All studies agree, however, that by the time these stars reach the main
sequence their protoplanetary discs must have been dispersed.
\textit{IRAS} showed that some 15\% of main sequence stars harbour dust
discs (Lagrange et al. 2000) and the very brightest of these have been
detected in the sub-mm (Zuckerman \& Becklin 1993; Holland et al. 1998;
Greaves et al. 1998; Sylvester et al. 2001; Sheret et al. in preparation).
However, these are the only stars $>10$ Myr for which discs have been
detected in the sub-mm.
Also, these discs are both much less massive, typically $< 0.1 M_\oplus$
(Holland et al. 1998), than their younger counterparts,
and cannot be remnants of the protoplanetary disc, since the lifetime of this
material is shorter than the age of the stars (e.g., Backman \& Paresce 1993).
In fact, these discs must be continually replenished, probably by the
collisional break-up of asteroids or comets in the systems (Wyatt \& Dent
2002) and are henceforth referred to as debris discs.

Our understanding of how a protoplanetary disc evolves into a debris disc
as its host star evolves onto the main sequence has been hindered by the
difficulty of finding an unbiased sample of stars of intermediate age.
For late-type stars ($<2M_\odot$) it is well-known that few such stars have
been identified.
Herbig (1978) noted that the T Tauri phase corresponds to just a small
fraction of a star's pre-main sequence lifetime and inferred that 
a significant number of Post T Tauri stars with ages of 10-100 Myr should
exist.
However, such Post T Tauri stars are notoriously difficult to find, since
they show few signs of youth such as active accretion and are no longer
associated with regions of active star formation.
The more massive ($>2M_\odot$) counterparts to T Tauri stars,
Herbig AeBe stars, on the other hand, are found at all stages of the
pre-main sequence and even up to 10 Myr after arrival on the main sequence
(Waters \& Waelkens 1998).
Thus the evolution from protoplanetary to debris disc for higher
mass stars might occur after the star arrives on the main sequence.

One method for identifying stars of intermediate age is as the secondaries of
binary systems with an O or B type primary (Murphy 1969; Gahm, Ahlin \&
Lindroos 1983), since in this case the primary must be younger
than 150 Myr, and so, assuming coevality, must the secondary.
The most comprehensive list of such systems was compiled by Lindroos (1986;
hereafter L86) and is comprised of 84 companions in 78 systems.
This is often referred to as the \textit{Lindroos sample}.
Disc masses have been estimated for several of these Lindroos stars
using sub-mm and mm observations (Jewitt 1994; Gahm et al. 1994; Ray et al.
1995).
These studies showed no detection of dust around stars older
than 10 Myr, thus supporting the hypothesis that protoplanetary
discs have been depleted by this time (Jewitt 1994).
However, the resulting constraints on dust mass, while below the level
of T Tauri and Herbig AeBe discs, did not reach down to debris disc
levels.
This left open the question of whether protoplanetary discs are rapidly
depleted down to, or even below debris disc levels, or whether there is a
slow decline of disc mass with age (Jewitt 1994).
It was the purpose of the sub-mm observations described in this paper to
set limits on the dust around a subset of the Lindroos sample approaching
the level of the brightest debris discs.

A description of our subset of the Lindroos sample is given in
section 2.
The sub-mm observations are described in section 3 and in section 4 we
describe the results of these observations.
In section 5 this sample is cross-correlated with the \textit{IRAS}
database to search for evidence of dust emission in the far-IR.
In section 6 we discuss these observations and use them to derive dust
masses for this sample and discuss the implications for the evolution
of protoplanetary discs.
Our conclusions are given in section 7.

\section{Sample}
\label{sec-sam}
The Lindroos sample is comprised of 84 physical companions located
2-60 arcsec from O and B type primaries (L86).
These were selected from the Washington Double Star catalogue
(Gahm et al. 1983) and then several tests performed to eliminate
optical rather than physical associations (Lindroos 1985, hereafter L85).
In this manner the sample has already been streamlined from an original
list of 290 stars.
The ages of the primaries, which are mostly main sequence stars, in all
these systems were estimated from Stromgren photometry to be $<150$ Myr
with half of this sample younger than 30 Myr (L85).
Assuming the same age for the secondaries, which have spectral types
in the range B2 to K5, 37 of these must still be contracting toward the
main sequence with the remainder having recently arrived there (L85).

For our sample we chose stars from table 1 of L86 according
to the following criteria.
First that they have declinations $\delta > -50^\circ$.
Second, we chose only those systems with later ($>$B5) type primaries.
This is because the circumsecondary discs in such systems could have
been affected by their interaction with the strong stellar wind and
high photoionisation flux of the very luminous primary star
(e.g., Johnstone, Hollenbach \& Bally 1998; O'Dell 2001).
We also chose only systems with projected separations $> 300$ AU.
Jensen, Mathieu \& Fuller (1996) used sub-mm/mm observations of pre main
sequence binaries to show that disc masses are significantly lower
for those systems with separations between 1-100 AU, while the disc
masses in systems with separations $>100$ AU are indistinguishable from
those of single stars.
This implies that the evolution of protoplanetary discs in systems
with intermediate separation are considerably affected by the presence
of a companion, presumably because of the truncation of these discs
by gravitational perturbations.
Our constraints allow us to consider the evolution of the companions
and of their discs as being similar to that of single stars
independent of the evolution of the primary.

Finally, since the studies of Lindroos and co-authors, the physical
association of some of the Lindroos sample has been tested by other
authors.
One of the methods used was to search for indications of youth
in the secondary.
Such indications include:
high Li abundance (Pallavicini, Pasquini \& Randich 1992;
Mart\'{i}n, Magazz\`{u} \& Rebolo 1992),
detection of H$_\alpha$ or Ca H \& K emission (Pallavicini et al. 1992),
strong X ray emission (Hu\'{e}lamo et al. 2000; Hu\'{e}lamo et al. 2001),
similar space motion to other groups of young stars
(Mart\'{i}n et al. 1992).
Another method was to check for similarity in the radial velocities of
the two components (Mart\'{i}n et al. 1992).
In rejecting erroneous Lindroos stars from our sample we took all of
these studies into account, in particular omitting all pairs
designated as \textit{likely optical} by Pallavicini et al. (1992).

This process left us with 22 stars, which we further split into two categories
according to spectral type:
a group of 13 young low mass (YLM) stars ($>$F0) comprising Post T Tauri
(PTT) and Young Main Sequence (YMS) stars;
and another of 9 young high mass (YHM) stars ($<$A9) comprising Post
Herbig AeBe (PHAeBe) stars and YMS stars.
The characteristics of these stars are shown in Table 1.
Of the group of YLM stars, 7 were identified by L85 as still
contracting toward the main sequence, while they identified just one of the
YHM stars (HD47247B) as not having yet reached the main sequence.
The lower fraction of pre-main sequence stars among the higher mass stars
in this sample is inevitable given their shorter contraction time
relative to the main sequence lifetime of the B-type primaries (L85).

It should be pointed out that the studies described above to test
for physical association of the Lindroos sample dealt only with
potential members of our low mass group.
For this reason we should anticipate that some fraction of our
YHM group could in fact be optical pairs (perhaps even up to
50\%, Pallavicini et al. 1992).
Also, since this list was compiled, a study was published which
compared the ages of the two components of 10 of our YLM sample
based on their position in the HR diagram relative to evolutionary
tracks (Gerbaldi, Faraggiana \& Balin 2001).
These authors classified three of the stars in our YLM sample
(HD90972B, HD108767B and HD127304B) as young, $\leq 100$ Myr, a
finding corroborated by the detection of X ray emission toward
these stars (Hu\'{e}lamo et al. 2000), but not associated with the
putative primaries, since the ages they derived for the three
primaries were all $200-240$ Myr.
Evidently there is still some uncertainty in resolving whether
the Lindroos stars are physically associated as well as in
determining their ages (see e.g., Hubrig et al. 2001 for another
estimate of the ages of 3 of our YLM primaries).
In the following discussion we adopted the ages given in L86
where available.


\begin{table*}
\begin{minipage}{180mm}
  \begin{center}
    \caption{Properties of the Lindroos stars in our sample.
     Spectral types are taken from L86.
     Ages are also from L86 except for the one indicated by an asterisk
     (HD77484);
     since L86 did not provide an age estimate for this star,
     its age was taken from Gerbaldi et al. (2001).
     Distances are from Hipparcos except those (indicated by an asterisk)
     for which this was not determined with $> 3\sigma$ uncertainty;
     distances to those stars were taken from L86.
     Projected separations are from the Washington Double Star catalog
     (Worley \& Douglass 1996), and the location of the secondary (or in one case
     tertiary) has been calculated using these offsets from the J2000 position of
     the primary from SIMBAD.
     The orbital semimajor axes ($a$ in arcsec) of these binary systems have 
     been estimated from their most statistically likely values based on the 
     observed separation ($\rho$ in arcsec): $\log{a} = \log{\rho} + 0.13$ 
     (Duquennoy \& Mayor 1991).
     Objects for which excess emission is reported in the current work are 
     shown in bold.}
    \begin{tabular}{lccclcccccc}
        \hline
        Primary &  & & & Companion & & & & & & \\
        Name & Sp & Age, Myr & Dist, pc & Name & Sp & Sep, " & $a$, AU & Sep, PA & RA(J2000) & Dec(J2000) \\
        \hline
        \multicolumn{3}{l}{(a) Low Mass Companions} &&&&&&&&\\
        HD560    & B9V    & $<50$  & 100    & HD560B             & G5VE & 7.7   & 1000   & $160^\circ$ & 00h10m02.38s & $+11^\circ$08'37.7" \\
\textbf{HD1438}  & B8V    & 95     & 212    & \textbf{HD1438B}   & F3V  & 6.2   & 1800  & $240^\circ$ & 00h18m41.67s & $+43^\circ$47'25.0" \\
        HD17543  & B6IV   & 62     & 185    & HD17543C           & F8V  & 25.2  & 6300  & $110^\circ$ & 02h49m19.21s & $+17^\circ$27'42.9" \\
        HD27638  & B9V    & 123    & 82     & HD27638B           & G2V  & 19.5  & 2200  & $25^\circ$  & 04h22m35.55s & $+25^\circ$38'03.2" \\
        HD33802  & B8V    & 40     & 74     & HD33802B           & G8V  & 12.7  & 1300  & $337^\circ$ & 05h12m17.56s & $-11^\circ$51'57.5" \\
        HD77484  & B9.5V  & 170(*) & 250    & HD77484B           & G5V  & 4.8   & 1600  & $92^\circ$  & 09h02m50.97s & $+00^\circ$24'29.3" \\
        HD90972  & B9.5V  & 120    & 147    & HD90972B           & F9VE & 11.0  & 2200  & $226^\circ$ & 10h29m34.77s & $-30^\circ$36'33.0" \\
        HD108767 & B9.5V  & $<112$ & 27     & HD108767B          & K2VE & 24.1  & 880   & $214^\circ$ & 12h29m50.92s & $-16^\circ$31'15.6" \\
\textbf{HD112413}& A0IIIP & $<28$  & 34     & \textbf{HD112412}  & F0V  & 19.3  & 890   & $228^\circ$ & 12h56m00.45s & $+38^\circ$18'53.3" \\
        HD127304 & A0V    & $<79$  & 107    & HD127304B          & K1V  & 25.8  & 3700  & $256^\circ$ & 14h29m47.71s & $+31^\circ$47'22.1" \\
        HD129791 & B9.5V  & 45     & 130    & HD129791B          & K5V  & 35.3  & 6200  & $206^\circ$ & 14h45m56.20s & $-44^\circ$52'34.8" \\
        HD143939 & B9III  & $<32$  & 168    & HD143939B          & K3V  & 8.6   & 1900  & $217^\circ$ & 16h04m44.04s & $-39^\circ$26'11.7" \\
        HD145483 & B9V    & $<71$  & 91     & HD145483B          & F3V  & 3.8   & 470   & $71^\circ$  & 16h12m16.31s & $-28^\circ$25'01.1" \\
        \multicolumn{3}{l}{(b) High Mass Companions} &&&&&&&&\\
        HD3369   & B5V    & 56     & 163(*) & HD3369B            & A6V  & 35.9  & 7900  & $173^\circ$ & 00h36m53.20s & $+33^\circ$42'34.0" \\
        HD35173  & B5V    & 44     & 331(*) & HD35173B           & B7V  & 26.0  & 12000 & $285^\circ$ & 05h23m30.00s & $+16^\circ$02'32.5" \\
        HD47247  & B5V    & 14     & 230    & HD47247B           & A2V  & 9.1   & 2800  & $336^\circ$ & 06h36m40.79s & $-22^\circ$36'44.7" \\
        HD63065  & B9.5V  & 110    & 293(*) & HD63065B           & A2V  & 17.4  & 6900  & $9^\circ$   & 07h47m02.65s & $+00^\circ$01'23.3" \\
\textbf{HD74067} & B9V    & 63     & 86     & \textbf{HD74067B}  & A2V  & 4.0   & 460   & $68^\circ$  & 08h40m19.49s & $-40^\circ$15'48.4" \\
        HD91590  & AP     & 38     & 259(*) & HD91590B           & AP   & 28.4  & 9900  & $162^\circ$ & 10h33m33.27s & $-46^\circ$59'00.2" \\
\textbf{HD99803} & B9Vp   & 126    & 100(*) & \textbf{HD99803B}  & A3V  & 13.1  & 1800  & $168^\circ$ & 11h28m35.33s & $-42^\circ$40'39.9" \\
        HD159574 & B81b   & 46     & 938(*) & HD159574B          & B7V  & 12.9  & 16000 & $341^\circ$ & 17h37m19.37s & $-40^\circ$19'00.2" \\
        HD177817 & B8IV   & 100    & 224(*) & HD177817B          & A0V  & 6.4   & 1900  & $2^\circ$   & 19h06m52.14s & $-16^\circ$13'39.0" \\
       \hline
    \end{tabular}
    \label{tab:tab1}
  \end{center}
\end{minipage}
\end{table*}

\section{Observations}
The observations were made using the Submillimetre Common-User
Bolometer Array, SCUBA (Holland et al. 1999) at the James Clerk
Maxwell telescope (JCMT).
Photometry observations were made at 850 $\mu$m wavelength,
using the central bolometer on the long-wave array.
The locations of the 22 observed (secondary) stars are given in
Table 1;
the pointing accuracy of the JCMT is estimated to be $\pm 2$ arcsec
rms.
Two of our sources were observed at 5-6 arcsec from these locations,
in one instance because we used coordinates for the secondary taken
from SIMBAD rather than calculated from the offset from the location
of the primary (HD33802), and the other because of a coordinate
transcription error (HD127304).
Since the instrumental beam size for these observations is 14.5
arcsec, we would expect to detect the flux from a point source 
at 60-70\% of its peak value;
the dust masses derived from the observed fluxes given in Table 2
have been scaled accordingly.
Each of our 22 sources was observed for $2 \pm 1$ hours during observing
runs spread out over the period February 2001 to January 2002.

The conventional two position chopping and nodding technique
was employed to remove the dominant sky background using a 60 arcsec
chop throw in azimuth.
The data were corrected for atmospheric extinction using sky opacities
that were measured at the JCMT using the skydip method at appropriate
intervals throughout the nights (Archibald et al. 2002).
The data for each night were calibrated using Mars or Uranus
when available, otherwise using the standard secondary JCMT calibrators.
The data reduction was accomplished using the \textit{SURF} package
(Jenness \& Lightfoot 1998), and anomalous signals were clipped
above the 3 $\sigma$ level.

SCUBA photometry observations also result in data for a further
36 bolometers on the long wave array.
The arrangement of these bolometers is such that they cover
a $\sim 2.3$ arcmin diameter field of view,
however, they are spaced so that the sky is instantaneously
undersampled, thus this data cannot be used to recreate a fully
sampled map.
The closest of these bolometers is 21 arcsec from the target,
corresponding to $\sim 600$ AU from our most nearby target.
Thus we did not expect any circumstellar emission to appear
in the remaining bolometers and their data was used to
remove sky level variations.
However, in section 4.2 we also used these bolometers to 
test for the presence of more extended emission around these
stars.
In this instance the bolometers were weighted according to their
noise and just the outer $\sim 70$ arcsec ring of bolometers
was used for sky removal.

\section{Results of SCUBA Observations}

\subsection{Photometry}
The results of the observations described in section 2 are given
in Table 2.
The 1 $\sigma$ uncertainty achieved in these observations
is $2 \pm 1$ mJy, resulting in 3 $\sigma$ upper limits of
$\sim 5$ mJy.
This is an order of magnitude improvement over previous
observations which were obtained at the JCMT using the
instrument UKT14 (Jewitt 1994).
Sub-mm emission was detected in the central bolometer toward
three stars in our sample at the 3 $\sigma$
level, HD74067B ($7.2 \pm 1.5$ mJy), HD112412 ($3.8 \pm 1.1$ mJy)
and HD99803B ($4.7 \pm 1.6$ mJy).
Two of these stars are from the high mass group, resulting
in a detection rate of 22\% for this group, while one of the
low mass stars was detected (a detection rate of 8\%).

\begin{table*}
\begin{minipage}{160mm}
  \begin{center}
    \caption{Properties of the Lindroos stars in our
    sample both observed and derived in this paper.
    All fluxes are in mJy.
    \textit{IRAS} fluxes come directly from the Faint or Point
    Source Catalogs (FSC and PSC respectively),
    apart from those indicated by an asterisk which were calculated
    using SCANPI (see text for details).
    \textit{IRAS} data has not been colour corrected.
    Upper limits from the PSC and SCANPI are quoted at the 3 $\sigma$
    level, while those from the FSC are at the 90\% confidence level.
    Disc masses only include the dust mass and were calculated using
    Equation 1;
    upper limits correspond to the 3 $\sigma$ uncertainty and assume
    a dust temperature of 30 K, while the range of masses for detected
    discs correspond to the range of possible disc temperatures
    (see text and Table 3).}
    \begin{tabular}{lccccccc}
        \hline
        Source & IRAS & $F_{12}$ & $F_{25}$ &
        $F_{60}$ & $F_{100}$ & $F_{850}$ & Disc mass, $M_\oplus$ \\
        \hline
        \multicolumn{3}{l}{(a) Low Mass Companions} &&&&&\\
        HD560B    & F00074+1051 & $278\pm44$ & $<245$ & $<130$ & $<1040$ &
          $-2.3 \pm 1.5$ & $<0.5$ \\
\textbf{HD1438B}  & F00160+4330 & $128\pm31$ & $206\pm27$ & $413\pm62$ & $<1370$ &
          $-2.4 \pm 1.7$ & $0.05$ \\
        HD17543C  & F02465+1715 & $269\pm27$ & $<137$ & $<307$ & $<1090$ &
          $1.2  \pm 1.6$ & $<1.8$ \\
        HD27638B  & F04195+2530 & $375\pm45$ & $<218$ & $<208$ & $<2010$ &
          $0.4  \pm 1.5$ & $<0.3$ \\
        HD33802B  & F05099-1155 & $508\pm30$ & $113\pm20$ & $<242$ & $<2720$ &
          $-1.2 \pm 1.7$ & $<0.5$ \\
        HD77484B  &             & & & & & $1.6  \pm 1.1$ & $<2.3$ \\
        HD90972B  & F10273-3021 & $252\pm30$ & $<85$ & $<269$ & $<1660$ &
          $0.6  \pm 1.6$ & $<1.1$ \\
        HD108767B & F12272-1614 & $2560\pm179$ & $649\pm71$ & $<217$ & $<497$ &
          $1.0  \pm 1.5$ & $<0.04$ \\
\textbf{HD112412} & F12536+3835 & $2340\pm140$ & $605\pm61$ & $110 \pm 31$ (*) & $<270$ (*) &
          $3.8  \pm 1.1$ & $0.02-0.06$ \\
        HD127304B & F14276+3200 & $153\pm23$ & $<77$ & $<210$ & $<379$ &
          $1.1  \pm 1.1$ & $<0.6$ \\
        HD129791B &             & & & & & $0.5  \pm 2.0$ & $<1.1$ \\
        HD143939B &             & & & & & $-3.3 \pm 1.9$ & $<1.8$ \\
        HD145483B & F16091-2817 & $298\pm36$ & $<184$ & $<632$ & $<2700$ &
          $-0.6 \pm 2.0$ & $<0.5$ \\
        \multicolumn{3}{l}{(b) High Mass Companions} &&&&&\\
        HD3369B   & F00341+3326 & $485\pm34$ & $<120$ & $<99$ & $<577$ &
          $1.4  \pm 1.4$ & $<1.2$ \\
        HD35173B  &             & & & & & $-3.0 \pm 1.7$ & $<6.1$ \\
        HD47247B  & F06345-2234 & $120\pm26$ & $<70$ & $<120$ & $<742$ &
          $-1.5 \pm 1.6$ & $<2.8$ \\
        HD63065B  &             & & & & & $-1.1 \pm 1.4$ & $<4.0$ \\
\textbf{HD74067B} & P08385-4005 & $365\pm40$ & $<355$ & $<910$ (*) & $<10600$ (*) &
          $7.2  \pm 1.5$ & $0.3$ \\
        HD91590B  &             & & & & & $-0.7 \pm 2.3$ & $<5.1$ \\
\textbf{HD99803B} & F11261-4223 & $402\pm36$ & $<117$ & $<186$ & $<1370$ & 
          $4.7  \pm 1.6$ & $0.2-0.9$ \\
        HD159574B &             & & & & & $-1.3 \pm 2.9$ & $<84$ \\
        HD177817B &             & & & & & $1.0 \pm 1.8$ & $<3.0$ \\
       \hline
    \end{tabular}
    \label{tab:tab2}
  \end{center}
\end{minipage}
\end{table*}

Given the high sensitivity of the observations, we might expect
some detections in our sample simply from background sources
such as high redshift galaxies falling within our beam.
Several estimates have been made of sub-mm number counts,
the most recent being the SCUBA 8 mJy survey
(Scott et al. 2002).
Scott et al. estimate that there are about 620 sources
per square degree above 5 mJy at 850 $\mu$m;
as many as $\sim 2000$ sources per square degree should be
detected above 3 mJy (Eales et al. 2000).
Thus we would expect there to be one background source in any
given SCUBA field of view (2.3 square arcmin) at the level of
$> 5$ mJy.
However, only one in a hundred of these sources would fall within
the central photometry bolometer.
Thus the probability of erroneously detecting a 5 mJy source in
our sample of 22 sources is about 1 in 5.
While there is a small (20 \%) chance that one of our detections is
in fact of a background source, it is very unlikely that all
three sub-mm sources are unassociated with the program stars.

With such a high sensitivity it is also necessary to check
whether we might (or indeed should) have detected the
photospheric emission of any of the stars.
However, even the two closest primaries (HD108767 and HD112413)
have predicted photospheric fluxes of $< 0.6$ mJy at 850 $\mu$m.
The more distant primaries, as well as all the secondaries,
have predicted fluxes much less than this.

Another consideration for our observations given the large
beam size is that for several of our sources the primary falls
within the beam when pointing at the companion (e.g., if their
separation is less than about 7-8 arcsec, see Table 1).
In such cases any emission that is detected could originate near
the primary and not the secondary star.
Consider the HR4796 wide binary system which hosts a
circumprimary disc, but not a circumsecondary disc
(Jayawardhana et al. 1998).
At 7.7 arcsec separation a SCUBA 850 $\mu$m observation centred
on the companion would have detected $\sim 46$ \% of the emission from
the circumprimary disc (i.e., $\sim 9$ mJy Greaves, Mannings \&
Holland 2000).
This uncertainty is relevant for our detection of emission
toward HD74067B, since this star is just 4 arcsec from
HD70467A.
If the emission we detected for HD74067B is actually centred
on the primary (and point-like), it would have been underestimated
in Table 2 by 80\%.
The emission detected in the vicinity of HD112412 and HD99803B,
however, could not have its origin near the primary stars in these
systems, since at 19.3 and 13.1 arcsec offset, these lie outside
the beam.

\subsection{Search for Extended Emission}
In general any emission from circumstellar discs in the
systems we observed should come from within a few arcsec
of the stars assuming these discs have a typical size of less
than a few hundred AU, and so should lie within the central
bolometer.
However, in the case of HD74067 we were inspired to search
for emission from the region around this system because it
is in the galactic plane (galactic latitude $b=0.52^\circ$)
with extended cirrus emission nearby (though not peaked on
this system) evident from the \textit{IRAS} 100 $\mu$m image.
It was thus questionable as to whether the detection is of
a circumstellar disc, cirrus heated by the stars, or indeed
the chance alignment of a background or foreground cirrus
hotspot.
One test is to see if the emission is extended on greater
than arcminute scales, since this is a characteristic
of other cirrus hotspots detected both in the far-IR
from its thermal emission (Gaustad \& Van Buren 1993) and
in the optical as reflection nebulosity (Kalas et al. 2002).

Thus we repeated the data reduction using the outer ring
of 18 bolometers at 62-82 arcsec from the centre of the array
to estimate the sky level.
We then looked at the mean flux levels in the rings of 6
and 12 bolometers at 21-28 arcsec and 40-55 arcsec 
from the centre respectively;
this part of the reduction was done in IDL.
Significant emission was indeed detected in both rings,
decreasing from an average mJy/beam of $6.7 \pm 1.4$ 
at the centre\footnote{The average mJy/beam at the
centre is a factor of 1.075 lower than the mJy derived by
reducing photometry observations (which is the figure given
in Table 1), since during such observations the central bolometer
performs a 4 square arcsec 9 point jiggle pattern about the centre
(Holland et al. 1999).
Reduction of photometry accounts for this to find the flux at the
centre assuming we are observing a point source.}
to $2.5 \pm 0.6$ at 25 arcsec and $1.9 \pm 0.5$
at 47 arcsec from the star
The outer ring at 70 arcsec showed no emission ($0.3 \pm 0.4$
mJy/beam) as expected as this was used to estimate the sky level.

We also checked to see if this emission is symmetrically
distributed around the star.
To do this we put individual datapoints into different bins
according to their radial distance from the star and position
angle on the sky and then obtained the mean and standard deviation
in each bin with the different bolometers of different observations
weighted according to their noise.
In this way we split each bolometer ring into four quadrants.
The findings are summarised in Figure 1.
The emission does appear to be asymmetrical in that
the only significant emission component is in the N and S
directions for both bolometer rings.
Combining the data for diametrically opposite quadrants we find
that the average mJy/beam in the N+S direction at 25 and 47
arcsec is $4.2 \pm 0.9$ and $3.7 \pm 0.8$ respectively, while that
in the E+W direction is just $1.3 \pm 0.8$ and $0.4 \pm 0.7$
at the same distances.
We changed the position angle of the quadrant pattern in Figure 1
to maximise the asymmetry and estimate the error on this angle to
be $0 \pm 30^\circ$.
An asymmetry is also apparent in the bolometer ring at
70 arcsec, with an average mJy/beam of $1.5 \pm 0.6$ 
in the NE+SW direction compared with $-1.3 \pm 0.6$ NW+SE.
This indicates that the emission could extend to 70 arcsec
from the star.
If so, since the mean flux in this bolometer ring
was used to subtract the sky emission from the remaining
bolometers, the sky level would have been overestimated
and the fluxes given in Figure 1 are too low by $\sim 1.3$ mJy/beam.

\begin{figure}
  \begin{center}
    \begin{tabular}{rl}
      \textbf{(a)} & 
      \psfig{figure=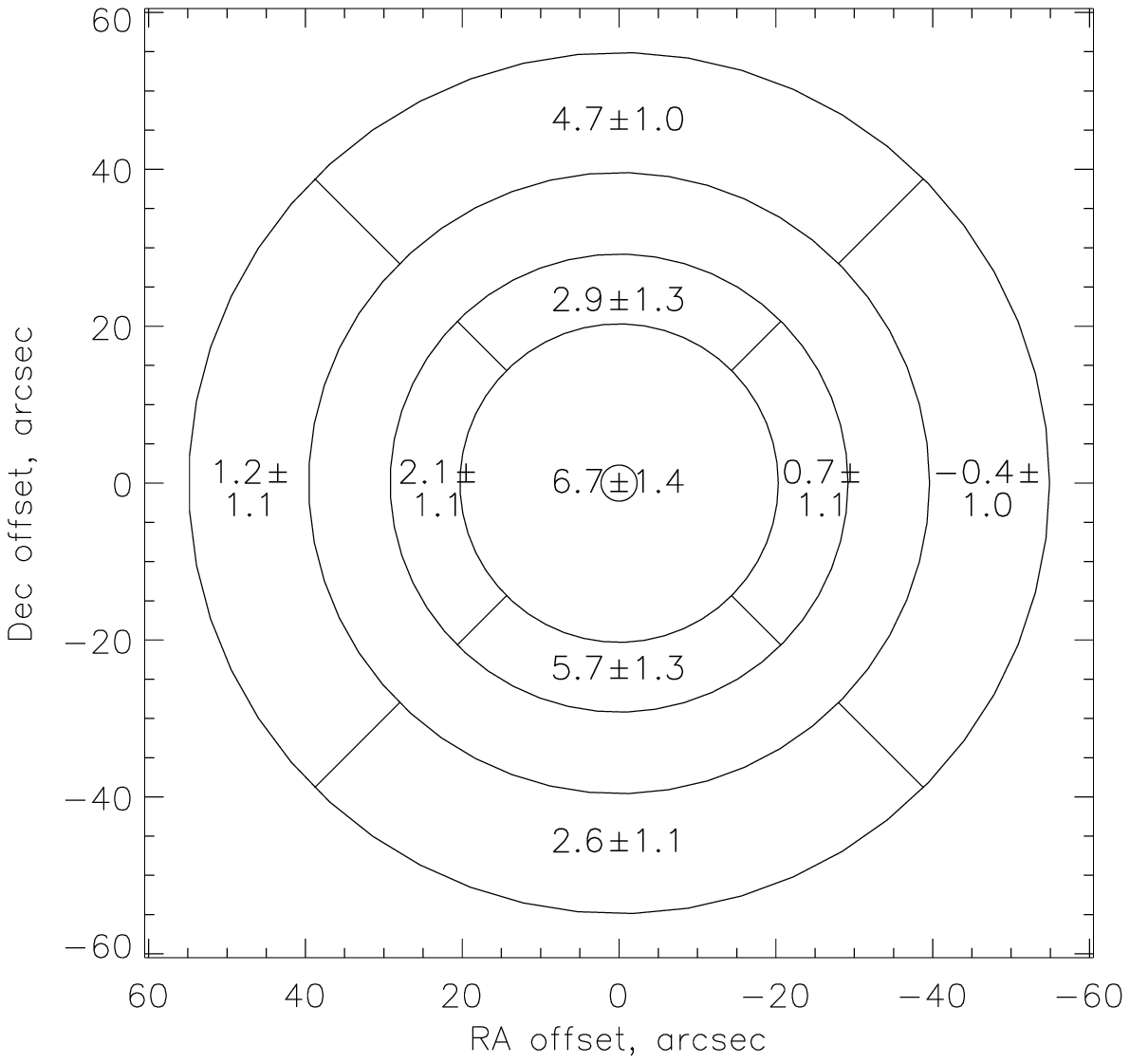,height=2.5in} \\
      \textbf{(b)} & \hspace{-0.25in}
      \psfig{figure=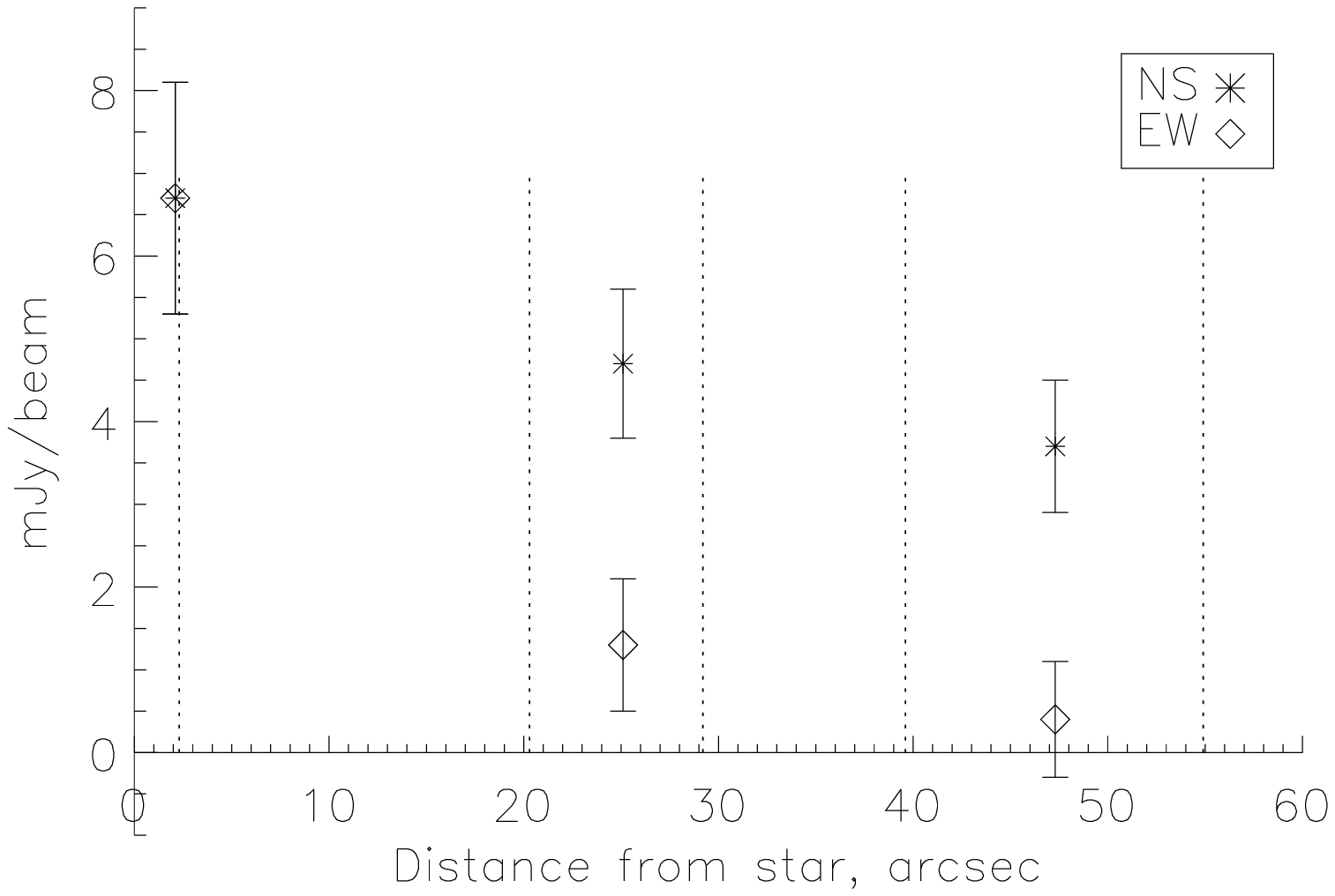,height=2.1in}
    \end{tabular}
    \caption{The distribution of the 850 $\mu$m emission
    detected around HD74067B:
    \textbf{(a)} Undersampled map.
    The circles show the extent of the first and second rings of
    bolometers at 21-28 arcsec and 40-55 arcsec respectively.
    These rings have been subdivided into quadrants and the
    mean (and its error) of the datapoints falling 
    into each quadrant are shown in units of the average
    mJy/beam in this region.
    \textbf{(b)} Radial distribution of emission.
    Each datapoint corresponds to the average of all data
    with a position angle within $90^\circ$ of N or S
    (asterisk symbol) and E or W (diamond symbol).}
    \label{fig:hd74067figs}
  \end{center}
\end{figure}

While it is possible that the asymmetry is caused by real
structure in the emitting material, the signals measured are
also consistent with a symmetrical centrally peaked structure
extending to $\sim 70$ arcsec from the stars.
In such an interpretation, the asymmetry arises because
some of the symmetrical emission has been artificially eliminated
by chopping just 60 arcsec onto the same structure.
Because the observations were all undertaken just before
or just after transit, the position angle of the azimuth chop was
within $\pm 40^\circ$ of E-W for all of our observations.
Thus if the emission is centrally peaked, our E+W observations 
should be lower than those N+S.
Assuming the observed N+S emission in Figure 1b is the level
at which E+W emission should have been observed if chopping
off-source, chopping on-source would have resulted in
E+W fluxes $\sim 3$ mJy/beam lower than those in the N+S
direction in both rings of bolometers, consistent with that observed;
the emission in the central bolometer would also have been
underestimated by a similar amount.
To test whether the structure is dependent on chop
position angle, we split the dataset into that taken before
and after transit, with chop position angles of
$72 \pm 7^\circ$ and $126 \pm 6^\circ$ respectively.
The position angle of the structure derived was indeed found
to be higher for the post-transit observations by
$\sim 45^\circ$ for all rings of bolometers.
We conclude that this emission extends to 70 arcsec from the
star, is centrally peaked and that a significant
component is symmetrically distributed around the star.

This process was now repeated for the other stars for which
emission was detected.
No significant extended emission was detected in the vicinity of
either HD99803B or HD1438B.
However, additional emission was detected in the bolometer
ring 25 arcsec from HD112412 where the average mJy/beam is
$2.0 \pm 0.4$.
Averaging the data points lying within 7.3 arcsec (i.e., within
one beam) of different points close to and within this ring
showed that this emission can be resolved into three distinct
sources (see Figure 2) with no significant emission detected in
the remainder of the ring:
(i) The first source appears to be centred $\sim 4$ arcsec N of
the primary HD112413 (i.e., at 14 arcsec E and 17 arcsec N of
HD112412) with a flux of $5.5 \pm 1.6$ mJy.
Since this source is undersampled, we estimate the uncertainty
in its position be $\pm 5$ arcsec and we attribute this emission
to a circumprimary disc;
this undersampling also means that the absolute level of the
emission is not well constrained, with an additional uncertainty
estimated to be $\sim 30$\%.
(ii) Another source is located 9 arcsec W and 29 arcsec N
of HD112412 with a flux of $9.5 \pm 1.9$ mJy
(with similar uncertainties in these values).
(iii) A third source is detected almost diametrically
opposite HD112413 at a location of 12 arcsec W and 15 arcsec
S of HD112412 with a flux of $5.7 \pm 1.9$ mJy.
We note that we would not be able to tell if the emission detected
toward HD112412, HD112413 and source (iii) forms part of an
extended structure, since the regions between the sources were
not sampled (see e.g., Figure 2).

\begin{figure}
  \begin{center}
    \begin{tabular}{c}
      \psfig{figure=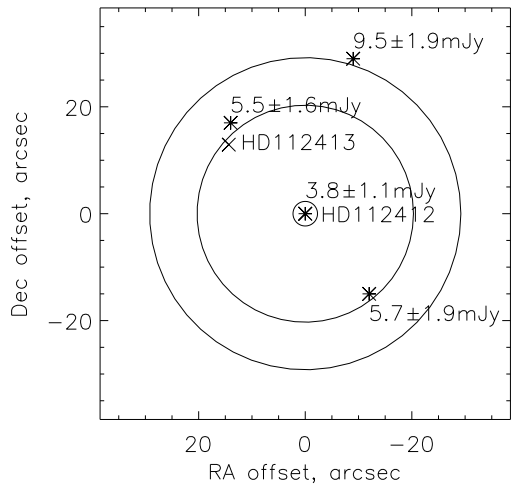,height=2.5in} \\
    \end{tabular}
    \caption{The distribution of the 850 $\mu$m emission
    detected around HD112412.
    The circles show the extent of the first ring of
    bolometers at 21-28 arcsec.
    The circumsecondary disc located at the centre
    of the map and the three additional sources detected
    in the first bolometer ring are shown by asterisks and
    their flux level is also given.
    The location of HD112413 is shown with a cross.}
    \label{fig:hd112412fig}
  \end{center}
\end{figure}

Due to the high proportion of detections we found using
this method (2/4), it was questioned whether the non-standard
reduction procedure was producing false results.
We looked for emission near all the remaining stars in our
sample.
The only detection was of emission of $7.4 \pm 2.1$ mJy
located 42 arcsec E and 31 arcsec S of HD145483B.
In the absence of fully sampled maps it is not possible to
determine the true nature of the offset sources (ii) and
(iii) near HD112412 and of the source near HD145483B.
However as the two additional bolometer rings we checked
cover a $\sim 5000$ square arcsec field of view, we would
expect to detect $\sim 5$ unrelated background sources
in our sample of 22 stars at the $\geq 5$ mJy level
(see section 4.1), thus we do not discuss these detections
further in this paper.

\section{Cross-correlation with \textit{IRAS}}
Also shown in Table 2 is a cross-correlation of our sample
with the \textit{IRAS} Faint Source and Point Source Catalogs
(hereafter FSC and PSC respectively).
These catalogs provide flux densities measured in four
wavelength bands centred at 12, 25, 60 and 100 $\mu$m,
and were searched for sources within 60 arcsec of our program
stars.
In this way emission was detected toward 14/22 of our program
stars.
Since the FSC is $\sim 2.5$ times more sensitive than the PSC,
these results are shown preferentially.
Just one of our detections (HD74067B) did not turn up in the
FSC, and this is because the FSC excludes sources within
$10^\circ$ of the galactic plane (i.e., those with
$|b| < 10^\circ$).

For all of these detections, the 12 $\mu$m flux is consistent
with that expected from the photospheres of the primary stars
in these systems by extrapolating their K (where available
in the literature or from the 2MASS database) or V band
magnitudes.
The photospheres of the three closest stars (HD108767, HD112413
and HD33802) were also detected at 25 $\mu$m.
For one of the stars, HD1438, excess emission was detected above
photospheric levels at both 25 and 60 $\mu$m at $185 \pm 31$ 
and $409 \pm 62$ mJy respectively\footnote{The colour corrected
photospheric emission has been subtracted from these fluxes, but
no colour correction applied to this excess}.
The positional uncertainty of the \textit{IRAS} source is such that we
cannot distinguish if this emission originates nearer the primary
or secondary component in this system.
This uncertainty is particularly pertinent when we consider that
a system which just missed inclusion in Lindroos' list because
the primary is a main sequence A0 star, HR4796, shows
evidence from mid-IR imaging of a dust disc around the primary but
not one around the pre-main sequence M-type secondary (Jayawardhana
et al. 1998).
At 7.7 arcsec separation, it had not previously been possible
to tell which star hosted the far-IR emission detected by
\textit{IRAS}.
We suggest that subsequent mid-IR imaging of this system
would resolve this issue.

A similar cross-correlation of the Lindroos sample with the
\textit{IRAS} catalogs was performed by
Ray et al. (1995) and included 10 of our low mass sample.
These authors reported the detection of non-photospheric emission
toward 7 of these sources in all four wavebands.
It appears that they interpreted all fluxes reported in the
\textit{IRAS} catalogs as positive detections, whereas the correct
interpretation of a flux reported with FQUAL=1 is that
it is an upper limit ($3\sigma$ for the PSC, 90\% confidence
for the FSC) as reported in this paper.

Even greater sensitivity can be obtained from the \textit{IRAS} database
using SCANPI (\textit{IRAS} Scan Processing and Integration), available
at the IRSA (Infrared Science Archive) website 
(http://irsa.ipac.caltech.edu).
This performs averaging of the \textit{IRAS} raw survey data scans 
passing close to a given location.
We used this to coadd the scans near our survey sources.
This resulted in better ($3\sigma$) upper limits to the far-IR fluxes at
the location of some of our sub-mm detections, as well as the
detection of the photosphere of HD112413 at 60 $\mu$m.
Also, emission was detected centred on HD17543C in the in-scan
direction ($342^\circ$ position angle) at 60 $\mu$m
with a peak of $220 \pm 23$ mJy.
However, this emission appears extended in the cross-scan
direction, since a similar level of emission appears
in scans taken up to 2 arcmin from the star.
Indeed there is a 60 $\mu$m only source of $247 \pm 42$ mJy
in the FSC, F02464+1714, located 106 arcsec from HD17543C
at a position angle of $254^\circ$.
We attribute this source either to local cirrus cloud heated
by the stars or to an unrelated object and do not discuss it
further in this paper.

\section{Discussion}

\subsection{SED Modelling and Dust Masses}
While it is clearly unrealistic to model the spectral energy
distributions (SEDs) of the infrared excesses of these stars
using just one or two datapoints, the upper limits obtained at
different wavelengths can set useful constraints on the
temperature of the emitting dust.
These can then be used to check whether the spatial extent
of the emission is consistent with that observed and to compare
the sensitivities of the \textit{IRAS} and SCUBA to the types of
dust emission that were detected.

The SED is modelled here assuming what is sometimes referred
to as \textit{modified black body} emission.
The dust is assumed to emit at a single temperature, $T$,
and its emission efficiency $Q_\nu$ is assumed to be unity
(i.e., to emit like a black body) below a critical wavelength,
$\lambda_0$, and to fall off $\propto \lambda^{-\beta}$ at
longer wavelengths.
Comparison with the observed and predicted emission
properties of real grains shows that $\lambda_0 \approx a$,
the typical size of the dust grains, and that $\beta$ lies
in the range 0.5-2 (Bohren \& Huffman 1983; Pollack et al. 1994).
Interstellar extinction curves show that interstellar-type
grains have $\beta \approx 2$ (e.g., Mathis 1990).
Modelling of the observed SEDs of young stellar objects
(YSOs), T Tauri discs and debris discs around main sequence stars
shows that at the early stages of a disc's evolution,
$\beta \approx 1.5$ while for more evolved discs
$\beta = 0.5-1$ (Dent, Matthews, \& Ward-Thompson 1998;
Dent et al. 2000).
Thus, given the age of our sample, we expect their disc
emission to be fitted by grains with $\beta$ close to 1.
In the following discussion we set $\beta = 1$ and
$\lambda_0 = 100$ $\mu$m, and for comparison show also
model fits for $\beta = 0$ and 2.
The results of the modelling for each star are discussed
below (see also Figure 3 and Table 3).

Dust masses were estimated from observed sub-mm flux,
since these give a reliable estimate of a disc's
dust mass (Zuckerman 2001), although for the far-IR-only
detection we used the 60 $\mu$m flux.
The appropriate formula is (e.g., Zuckerman \& Becklin
1993; Zuckerman 2001):
\begin{equation}
  M_{dust} = \frac{F_\nu D^2}{\kappa_\nu B_\nu(T)},
\end{equation}
where $F_\nu$ is the observed flux, $D$ is the distance to the
star, $\kappa_\nu = 0.75Q_\nu/a\rho$ is the dust opacity at
the observed wavelength ($\rho$ is the dust density), and
$B_\nu$ is the black body intensity at the dust temperature.
For consistency with the masses derived by other authors
(e.g., Zuckerman \& Becklin 1993; Holland et al. 1998; Greaves
et al. 1998; Sylvester, Dunkin, \& Barlow 2001) we adopted
a dust opacity of 0.17 m$^2$/kg at 850 $\mu$m, and scaled
to 60 $\mu$m assuming $\beta = 1$ (giving an opacity of
2.4 m$^2$/kg).
For dust temperatures of 30-100K, observed fluxes at 850 $\mu$m
correspond to $11-2.7 \times 10^{-3}D^2F_\nu$ Earth masses, where
$D$ is in parsec;
dust masses derived from 60 $\mu$m fluxes are much more dependent
on the assumed temperature of the dust.
The derived dust masses for our sample are reported in Table 2.

\begin{figure*}
\begin{minipage}{160mm}
  \begin{center}
    \begin{tabular}{rlrl}
      \textbf{(a)} & \hspace{-0.4in}
      \psfig{figure=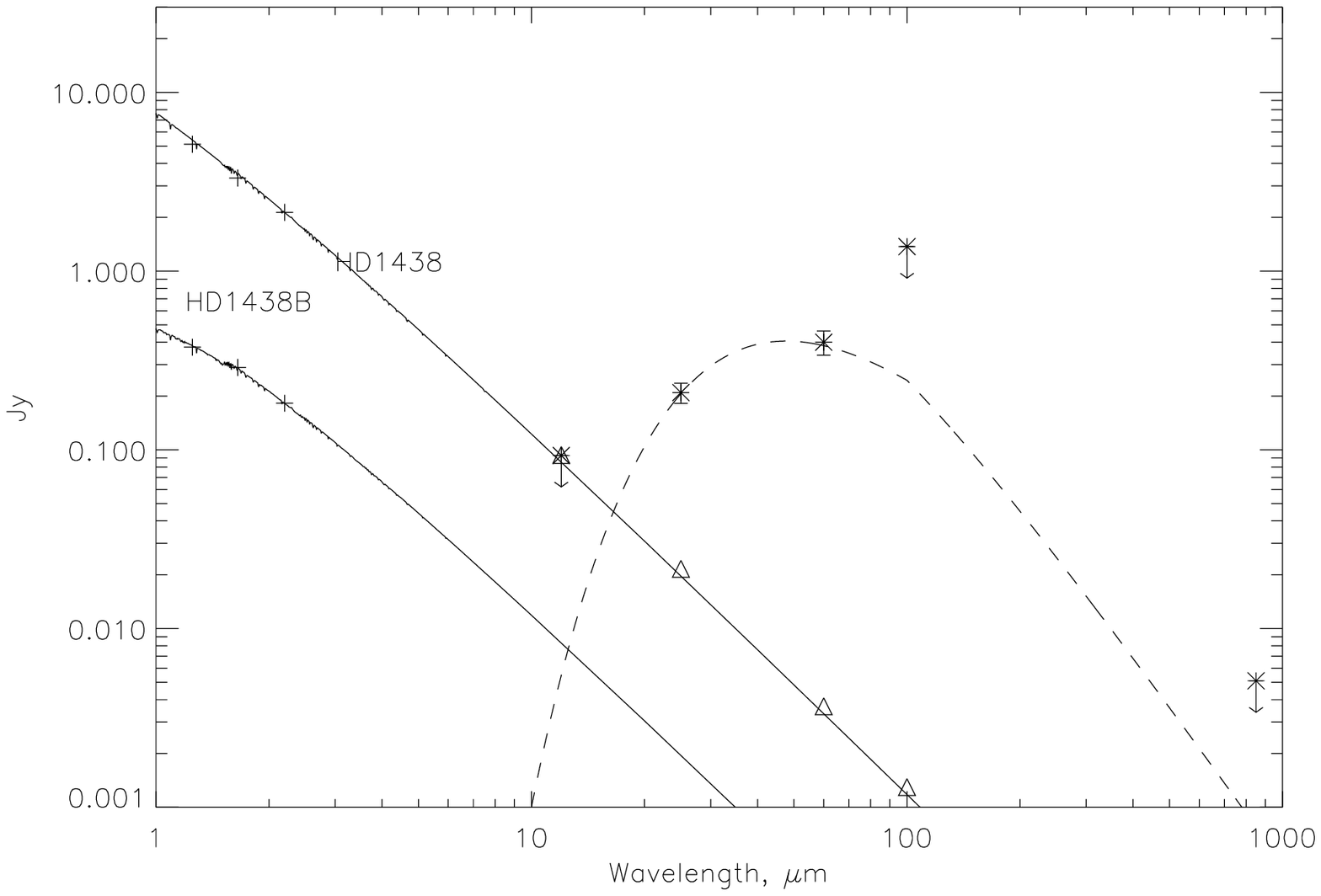,height=2.1in} &
      \textbf{(b)} & \hspace{-0.4in}
      \psfig{figure=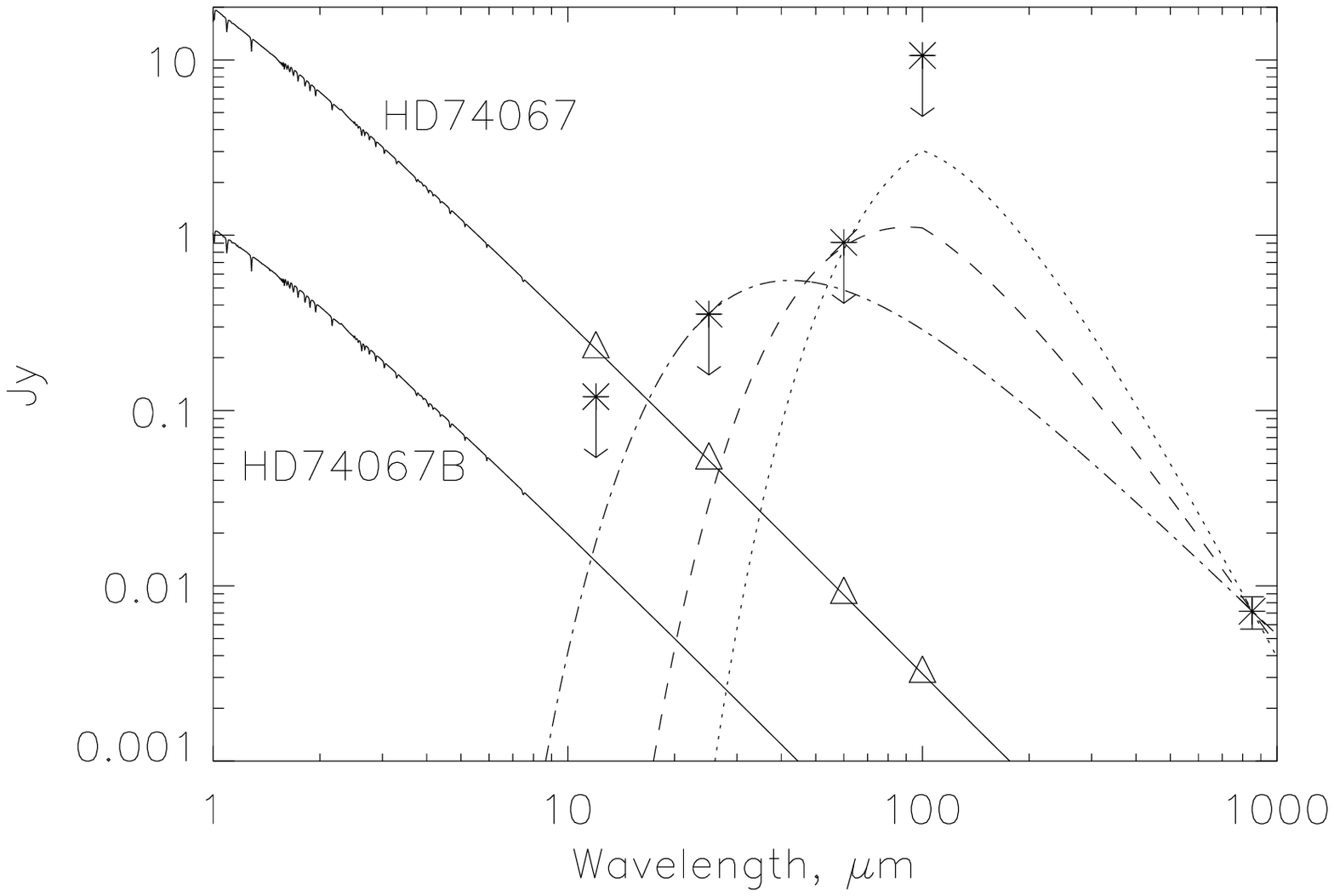,height=2.1in} \\
      \textbf{(c)} & \hspace{-0.4in}
      \psfig{figure=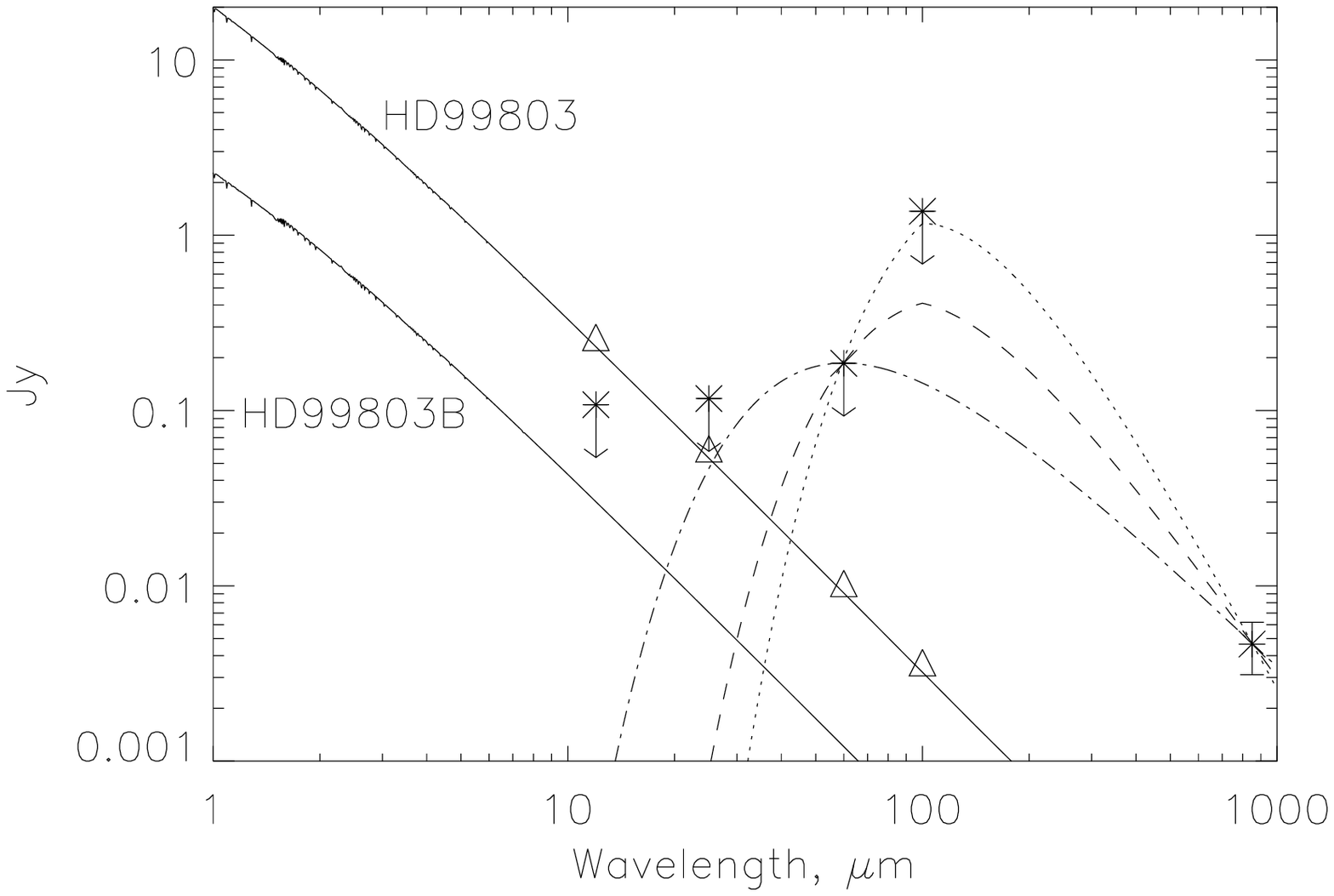,height=2.1in} &
      \textbf{(d)} & \hspace{-0.4in}
      \psfig{figure=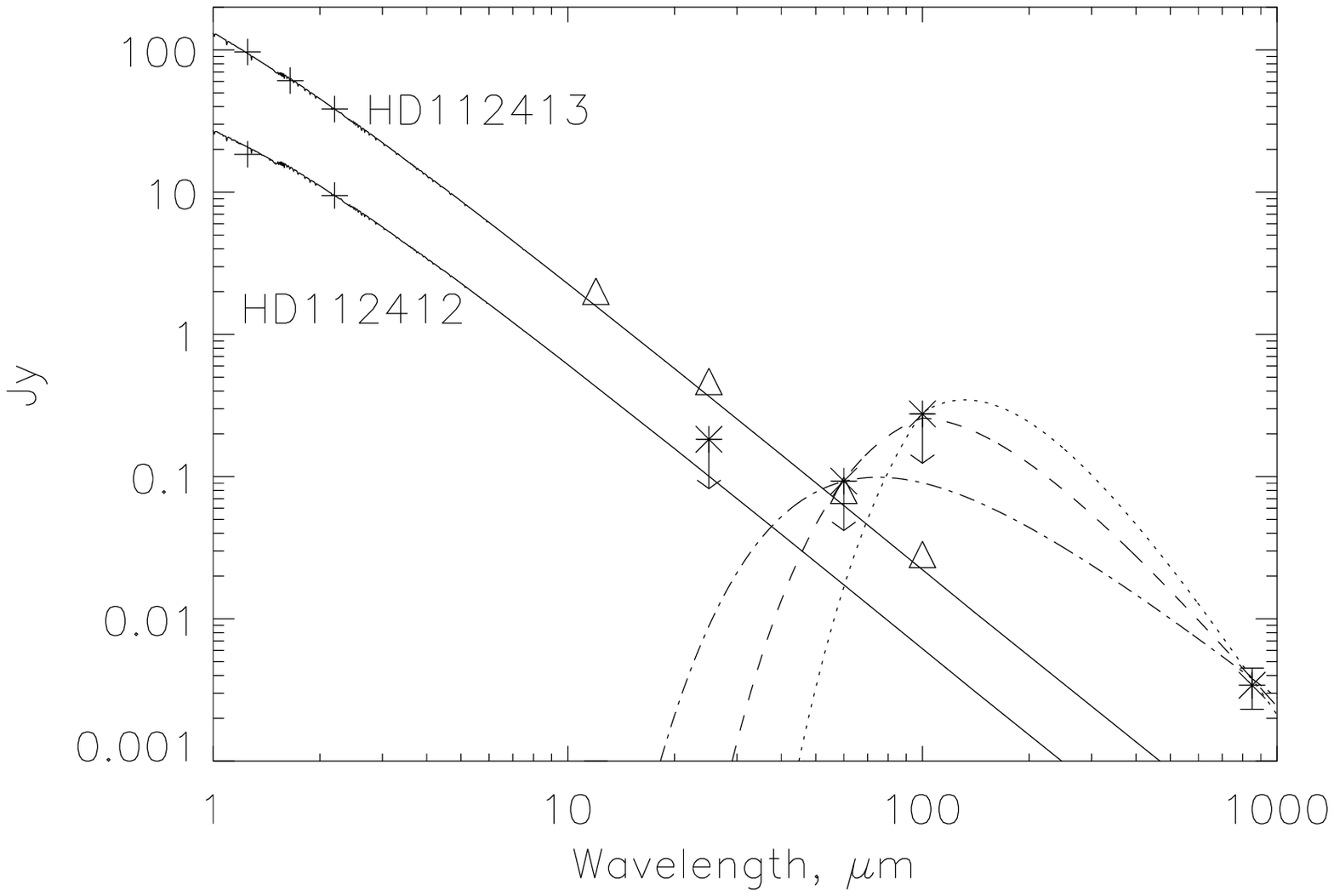,height=2.1in}
    \end{tabular}
    \caption{Spectral Energy Distributions of the emission
    toward the 4 stars for which excess emission is reported in
    this paper:
    \textbf{(a)} HD1438B, 
    \textbf{(b)} HD74067B,
    \textbf{(c)} HD99803B,
    and \textbf{(d)} HD112412.
    Stellar spectra were determined from Kurucz model
    atmospheres appropriate to the spectral type,
    and were scaled to the K magnitude (or V magnitude if
    unavailable) which are plotted with plus symbols.
    The \textit{IRAS} and SCUBA fluxes are plotted with asterisks.
    Since the \textit{IRAS} fluxes given in Table 2 include a contribution
    from the photospheric emission of both stars in the system
    (shown with triangular symbols), this has been colour
    corrected and subtracted before plotting.
    In the case of HD1438, the resulting fluxes have been
    further colour corrected for a 107 K black body.
    The dashed, dotted and dash-dotted lines show modified black body
    fits to the excess emission and are discussed in the text;
    note however that the fits in \textbf{(b)}, \textbf{(c)} and
    \textbf{(d)} are constrained by the textit{IRAS} upper limits
    and the actual far-IR emission could fall below this level.}
    \label{fig:seds}
  \end{center}
\end{minipage}
\end{figure*}

\begin{table*}
\begin{minipage}{160mm}
  \begin{center}
    \caption{Results of SED modelling for the emission 
    detected toward HD1438B, HD74067B, HD99803B, HD112412,
    (see Figure 3).
    Here we give the derived dust temperature $T$, the distance from
    the secondary star $r_B$ of black body grains emitting at this
    temperature, and the corresponding mass of the $M_{dust}$ for different
    assumed dust properties defined by the parameter $\beta$.
    The appropriate lines on Figure 3 for these models are also given.}
    \begin{tabular}{lccccc}
        \hline
        Source    & $\beta$      & $T$, K & $r_B$, AU          & $M_{dust}$, $M_\oplus$ & Figure 3        \\
        \hline
        HD1438B   & $>0.1$       & 107    & 10 (or $r_A=90$ AU)& 0.05                   & dashed          \\
        HD74067B  & 0.0          & $<120$ & $>30$              & $>0.1$                 & dash-dot        \\
                  & 1.0          & $<56$  & $>130$             & $>0.3$                 & dashed          \\
                  & 2.0          & $<34$  & $>340$             & $>0.5$                 & dotted          \\
        HD99803B  & 0.0          & $<85$  & $>50$              & $>0.2$                 & dash-dot        \\
                  & 1.0          & $<42$  & $>200$             & $>0.3$                 & dashed          \\
                  & 2.0          & $<29$  & $>420$             & $>0.5$                 & dotted          \\
                  &              & $>20$  & heating by primary & $<0.9$                 &                 \\
        HD112412  & 0.0          & $<68$  & $>40$              & $>0.02$                & dash-dot        \\
                  & 1.0          & $<38$  & $>140$             & $>0.04$                & dashed          \\
                  & 2.0          & $<22$  & $>410$             &  -                     & dotted          \\
                  &              & $>25$  & heating by primary & $<0.06$                &                 \\
        \hline
    \end{tabular}
    \label{tab:tab3}
  \end{center}
\end{minipage}
\end{table*}

\subsubsection{HD1438B}
Since the excess emission is detected at two wavelengths
we can derive its temperature to be $\sim 107$ K.
The modified black body fit is shown by the dashed line
in Figure 3a.
This temperature implies a distance of either 10 AU
from HD1438B or 90 AU from HD1438 if the dust emits
as a black body.
These distances would be larger if the grains are small
and so emit hotter than black bodies at an equivalent
distance (Wyatt et al. 1999).
Thus if this emission is circumprimary, it should 
be possible to resolve this disc in the mid-IR from an
8 metre class telescope.

It is not surprising that this disc was not detected in the
sub-mm, since the fit with $\beta = 1.0$ predicts
$F_{850} = 0.8$ mJy.
In fact our upper limit of 5.1 mJy at 850 $\mu$m only rules
out that the grains emit as perfect black bodies with
$\beta < 0.1$.
The upper limit to the dust mass derived from the sub-mm
flux is $0.6M_\oplus$, while the 60 $\mu$m flux implies
that we have detected material with a mass of just
$\sim 0.05M_\oplus$.
This is similar to the mass of the debris discs found around
young main sequence stars such as $\beta$ Pictoris (e.g.,
Holland et al. 1998).
We have also determined the fractional luminosity of the
dust disc with respect to that of the star,
$f = L_{ir}/L_\star$.
If this is a circumprimary disc $f \approx 0.3 \times 10^{-3}$,
a value typical of other debris discs, whereas if
circumsecondary $f \approx 0.01$, which would make it one of
the most luminous debris discs.

\subsubsection{HD74067B}
Consider first of all the sub-mm emission detected in the
central bolometer.
Figure 3b shows fits scaled to the 850 $\mu$m flux assuming
$\beta = 0.0,1.0,2.0$ (dash-dot, dashed and dotted lines
respectively).
These grain properties and the constraints from the \textit{IRAS}
non-detections imply dust temperatures of less than
120, 56, 34 K, and dust masses of at least 0.1,
0.3 and 0.5 $M_\oplus$ respectively.
It is possible to rule out interstellar-type ($\beta=2.0$) grains
as the origin of this emission, since such grains must lie
within 7.3 arcsec of HD74067B and, given the 4 arcsec projected
separation of this binary system, are also likely to be within
1000 AU of HD74067.
Such grains emit hotter than black bodies since they absorb
stellar radiation more efficiently than they re-radiate it
at longer wavelengths.
The emission efficiencies of interstellar-type grains
imply that they would be heated to $\sim 110 K$ at 1000 AU
from HD74067 (compared with a black body temperature of
$\sim 27$ K).
Since the observed grains are relatively cool for their
distance from the primary star (given the \textit{IRAS} limits),
we infer that they must have $\beta < 1.0$, although the exact
value depends on the assumed value of $\lambda_0$.

While the distribution and level of the emission detected $> 25$
arcsec from the star remains uncertain due to the nature of our
observing method, the fact that this emission was detected
implies a substantial mass of dust.
If the emission is symmetrically distributed about the
stars with a level of 4.2 and 3.7 mJy/beam at 25 and 47 arcsec,
this implies integrated fluxes of 45 and 75 mJy in rings of
one beam width at these distances.
Further assuming emission at black body temperatures
(19 and 14 K respectively based on the projected separations
from HD74067), this implies dust masses of 7 and
20 $M_\oplus$, with more mass expected to have remained
unobserved between the bolometer rings.

Similar arguments to those in the paragraph above argue against
the presence of interstellar-type grains, since these would be
heated to 84 and 66 K at 2200 and 4000 AU from HD74067.
However, we note that the interaction of the stars with an
interstellar dust cloud would mean that only the largest grains
from that cloud could penetrate close to the stars, since the
smaller (few 0.1 $\mu$m) grains which dominate the
interstellar extinction curves would be repelled by
radiation pressure (Artymowicz \& Clampin 1997).
Thus the cool temperature of the excess emission could not
on its own rule out the interaction of the stars with
a cirrus cloud as the cause of this excess.
However, the mean Galactic density of solid grains is
just $\sim 7 \times 10^{-15} M_\oplus$/AU$^3$
(Artymowicz \& Clampin 1997).
Thus, as we infer a mass density from our sub-mm observations
of at least $7 \times 10^{-11} M_\oplus$/AU$^3$, then if this dust
is interstellar in origin, the star must be in a region
of the galaxy with at least 10,000 times the mean Galactic
density.
This is unlikely, since at $\sim 60$ Myr this system should not
still reside in the cloud from which it was born, and
such a high density would only apply in molecular clouds
which fill just 0.2\% of the Galactic stellar disc.

We conclude that the material at $>25$ arcsec must
be either bound to HD74067 (e.g., in a massive $>27 M_\oplus$
extended $\sim 6000$ AU circumbinary disc or remnant protostellar
envelope) or an unrelated background object (e.g., a nearby galaxy
or Galactic cloud).
This system is unusual in our sample for two reasons which
argue in favour of the different interpretations:
the orbital semimajor axis of this pair (460 AU; Table 1)
is the smallest in the sample, thus increasing the chances of survival
of a circumbinary disc;
and this is the only system in the galactic plane, thus
increasing the chance of alignment with a background object.
Further observations of this region are required to
determine the origin of this emission.

We note that extended circumbinary envelopes have been proposed as
the replenishment mechanism of T Tauri discs in binary systems
(Prato \& Simon 1997), a proposal which may be supported by
interferometric observations of binary T Tauri systems which
detect a lower level of emission than single dish measurements,
possibly implying the presence of additional extended emission
(Jensen \& Akeson 2003).
Also, if this material is circumbinary, then
the orbital semimajor axis of the binary (460 AU; Table 1)
implies that any circumbinary disc would be truncated
within $\sim 9$ arcsec from a point roughly midway between
the two stars (see table 1 of Artymowicz \& Lubow 1994).
Thus, if this is the case, the emission detected in the central
bolometer would have to be circumstellar, not circumbinary, in
origin, and further would be truncated at $\sim 150$ AU
(1.7 arcsec) from either star (see table 1 of Papaloizou \&
Pringle 1977), again ruling out the models with $\beta>1.0$
(see Table 3).

\subsubsection{HD99803B}
The SED fit was scaled to the observed 850 $\mu$m flux, and upper
limits to the temperature of the dust emission calculated for
different $\beta$.
For $\beta = 1.0$, shown with the dashed line in Figure 3c,
this temperature must be below 42 K to avoid contradiction
with the non-detection by \textit{IRAS}.
This corresponds to dust at least 200 AU (2 arcsec) from
HD99803B, while the inferred orbital semimajor axis of this
system (1800 AU; Table 1) implies that any circumsecondary
disc should be truncated beyond 540 AU from HD99803B
(Papaloizou \& Pringle 1977).
This model implies a dust mass of at least $0.3 M_\oplus$
and a fractional luminosity of $0.6 \times 10^{-3}$.

Models with $\beta = 0.0$ and 2.0 are also shown on Figure 3c
with dash-dot and dotted lines, respectively.
These imply that if the dust emits as a black body, its
temperature could be as high as 85 K (dust as close as
50 AU with a mass of $0.2 M_\oplus$).
Interstellar-type dust with $\beta = 2.0$, however,
require that $T<29 K$ (dust as close as 420 AU with
a mass of $0.5 M_\oplus$).
Dust orbiting HD99803B would be heated to $>20$ K by the
1800 AU distant primary star giving an upper limit to the dust
mass of $0.9 M_\oplus$.

\subsubsection{HD112412 and HD112413}
Starting with HD112412, the SED model was scaled to the
observed 850 $\mu$m flux, from which upper limits to the
temperature of the dust emission were calculated for
different $\beta$.
For $\beta = 1.0$, shown with the dashed line in Figure 3d,
this temperature must be below 38 K to avoid contradiction
with the non-detection by \textit{IRAS}.
This corresponds to dust at least 140 AU (4 arcsec) from
HD112412 and implies a dust mass of at least $0.04 M_\oplus$
and a fractional luminosity of $4 \times 10^{-5}$.
Since the emission falls within the beam, we also know that
the emission arises $<250$ AU from HD112412.
This is consistent with the expected truncation radius
of the circumsecondary disc due to its tidal interaction
with the primary at 260 AU (Papaloizou \& Pringle 1977).

Models with $\beta = 0.0$ and 2.0 are also shown on Figure 3d
with dash-dot and dotted lines, respectively.
These imply that if the dust emits as a black body, its
temperature could be as high as 68 K (dust as close as
40 AU with a mass of $0.02M_\oplus$).
The dust cannot, however, have $\beta = 2.0$, as such models
require that $T<22 K$ and so that the dust is further than 410 AU 
from HD112412;
this constraint is incompatible with the emission
falling within the beam.
In fact both the temperature and distance constraints are also
inconsistent with the implied distance to the primary star of
890 AU (Table 1), since this would both truncate the circumsecondary
disc (see above) and heat dust falling in the beam centred on HD112412
to $>25$ K.
This 25 K lower limit to the circumsecondary dust temperature also
means that its dust mass is unlikely to be more than
$0.06 M_\oplus$.

Since the \textit{IRAS} upper limits also apply to the
emission from the disc around HD112413, and the circumprimary
850 $\mu$m emission is the same (within the uncertainty) as
that from the circumsecondary disc (see section 4.2), the
temperatures and masses derived from the SED models in Figure 3d
(see Table 3) are also valid for the circumprimary disc.
The factor of $\sim 8$ times higher luminosity of an A0III to
an F0V star, however, means that the limits from these models
to the distance of the dust from HD112413 should be approximately
2.8 times the value of $r_B$ given in Table 3.
Thus, since for this emission to fall within the beam it must
be $<250$ AU from the star, we can rule out models with $\beta$
greater than about 0.5;
i.e., to account for the \textit{IRAS} non-detections, the
circumprimary grains must emit very much like black bodies
at temperatures of $<49$ K.

\subsection{Detection Summary}
Of the 22 stars in our sample, we detected:
\begin{itemize}
  \item One warm (100-110 K) circumstellar disc toward HD1438
        with \textit{IRAS}, but not with SCUBA;
        it was not possible to tell if this is circumprimary
        or circumsecondary.
  \item One system, HD74067, with centrally peaked extended
        emission, possibly indicating the presence of
        a nearly face-on circumbinary disc/envelope extending to
        $\sim 6000$ AU, and an additional circumstellar component.
  \item Two cold ($<40$ K assuming $\beta=1$) circumsecondary discs
        around HD112412 and HD99803B in the sub-mm with SCUBA, but not
        in the far-IR with \textit{IRAS};
        the SCUBA observations also showed that one of these systems
        hosts a cold ($<50$ K assuming $\beta=0.5$)
        circumprimary disc (around HD112413).
\end{itemize}

A quick look at Table 1 shows that, at least within this sample,
there is nothing unusual about those systems with circumstellar
discs in terms of their age or binarity, since the detected
systems span the range of binary separation and age.
There does, however, appear to be a correlation with spectral
type, since all detections were around stars earlier than F3
($>3L_\odot$);
none of the eight solar-like G or K stars were found to harbour
discs.
There is also a correlation with distance in that the two
detections in the YHM group were of the closest stars of that
group, and that in the YLM group was of the second closest.
This leaves the possibility open that all stars in the YHM
group and the majority of stars in the YLM group have cold discs
that are similar in mass to those detected around HD99803B
and HD112412 (see Figure 4).

It is unusual that the circumprimary HD112413 disc is not
much more massive than the circumsecondary HD112412 disc,
since near-IR and mm observations of binary systems in
Taurus-Aurigae show that circumsecondary discs, if present,
are much less massive than circumprimary discs (White \& Ghez 2001;
Jensen \& Akeson 2003).
A more massive circumprimary disc is also expected from
theoretical predictions of binary star formation (Bate \&
Bonnell 1997), and is certainly true of the quasi-Lindroos
system HR4796 (Jayawardhana et al. 1998).
However, since HD112413 is the only system in our sample for
which information was obtained about its circumprimary emission,
we cannot draw any conclusions on this subject.

\subsection{Submillimetre Disc Population}
What is particularly striking about these results is that
9-14\% of this sample (2 or 3 out of 22) have discs that
were not detected by \textit{IRAS}.
This fraction is even higher at 20-30\% (2 or 3 out of 10) if we
consider just stars more massive than $1.5 M_\odot$ ($<$F0),
and even higher still (up to 4/11) if we include the the circumprimary
disc detected around HD112413.
To our knowledge, just one circumstellar disc, that around the
M1 star TWA7 (Zuckerman 2001), has been discovered
around an isolated star that was not detected first in the far-IR
by a space-based telescope such as \textit{IRAS} or \textit{ISO}.
The majority of searches for discs in the sub-mm have been
directed toward stars for which excess emission is already
known in the far-IR.
However, such searches are biased toward relatively warm discs.
The fact that SCUBA was able to detect these discs while \textit{IRAS}
could not must be because these discs are cold ($<40$ K if $\beta=1$;
section 6.1).
Thus these results imply that a significant population of
cold \textit{submillimetre discs} could be awaiting discovery in
more unbiased sub-mm surveys, a finding which is supported by the fact
that the TWA7 disc was discovered in a sub-mm survey of the TW
Hydrae Association (Zuckerman et al., in preparation), members of
which do not necessarily have \textit{IRAS} excesses.
Depending on their temperature, these cold discs 
could also be detected by deep far-IR observations
using, e.g., \textit{SIRTF} or \textit{SOFIA}, and indeed
such observations in conjunction with those in the sub-mm
are required to determine the temperature of these discs
\footnote{Note that until such observations are performed
the upper limits to the temperature of this cold disc
population are dependent on the assumptions about the
shape of the SED, and that a temperature of up to
$\sim 70$ K is possible if $\beta=0$.}.

Since current estimates put the fraction of main sequence
stars with discs at 15\% based on the discs that could be
detected by \textit{IRAS} (e.g., Plets \& Vynckier 1999), an
additional cold disc population (i.e., one that \textit{IRAS}
could not detect) with the same incidence rate found for our
Lindroos sample would mean that the true disc fraction could
have been underestimated by a factor of two.
However, it must be remembered that our sample is biased
toward young stars.
This may be particularly relevant, since TWA7 is also
young at $\sim 10$ Myr (Webb et al. 1999).
However, this may be observational bias, since out of the
three sub-mm surveys we know about that were undertaken
without selection toward stars with \textit{IRAS} excesses
(this work; Zuckerman et al., in preparation; Greaves et al., in
preparation) two were specifically directed toward young stars.

This does, however, raise the question of whether these discs
are \textbf{(a)} debris discs or \textbf{(b)} protoplanetary
remnants.
The distinction lies in whether these dust discs must be
continually replenished from the break-up of a population
of large planetesimals.
While the transition from protoplanetary to debris disc is
likely to be smooth, and there may be no definitive answer,
it is important to try to make the distinction because of
the implications for the incidence of these discs around
older stars.
If they can be shown to be protoplanetary remnants they would
be much less common around more evolved stars, whereas debris
discs could persist over the whole main sequence lifetime of
the parent star, albeit getting fainter or less common with
age (Habing et al. 1999; Spangler et al. 2001).
Also, if they can be shown to be debris discs, it implies
that significant grain growth has occurred at large distances
from the star.

\textbf{(a)} Planet formation models do predict the
existence of cold debris dust rings at large distances from the
star (Kenyon \& Bromley 2002).
In these models a collisional cascade is ignited at a given
distance from the star once planetesimals in this region
have grown to $\sim 1000$ km.
This makes a bright dust ring which subsequently decreases
in brightness until the supply of the 1 km planetesimals that
feed the cascade is exhausted.
Because of the longer planetesimal growth timescales 
at larger distances from the star, this means that as the
system evolves we would expect to see a bright dust ring
expanding out to larger radii until it reaches the edge of
the disc.
However, this model predicts that cold dust rings at $>150$ AU
would not occur until a few Gyr.
They also do not account for the effect of the binary companion
which would stir the planetesimal population making collisions
more energetic.
This could help the situation by igniting a collisional
cascade without having to wait for planetesimals to grow to
1000 km, although this may also hinder it by inhibiting
planetesimal growth in the first place.

\textbf{(b)} A different model predicts the existence of cold dust
rings left over from the protoplanetary disc (Clarke et
al. 2001).
This model was developed to explain the rapid removal on
a 0.1 Myr timescale of T Tauri discs at an age of
$\sim 10$ Myr, and shows how these
timescales can be achieved by the viscous evolution of a
gas disc in conjunction with its photoevaporation by
the central star.
In this model it is just the disc within the gravitational
radius ($\sim 7$ AU in their standard model) which is
dispersed rapidly (in 0.1 Myr) at 10 Myr;
the outer disc is subsequently dispersed over a $>10$ Myr
timescale as its inner edge expands outwards.
Thus Clarke et al. predicted the existence of cold discs
that would be preferentially detected in the sub-mm regime
beyond $\sim 28$ Myr, since by this time the only part of
the primordial gas disc that would remain would be outside
100 AU.
However, a more recent study claims that the rapid
clearing of the inner disc is not reproduced when
account is made for the reduction in photoionising flux
from the central source when the accretion flux is
reduced as the inner disc dissipates (Matsuyama, Johnstone
\& Hartmann 2003).
Rather this study predicts that the outer disc is
dispersed at the same rate as the inner disc.

While the models provide conflicting arguments as to the
likely origin of the dust, it may be possible to determine
this observationally.
Dust spirals in toward the central star due to the
Poynting-Robertson (P-R) drag force on timescales of
$400r^2/\alpha M_\star$ years, where $r$ is the distance of
the dust from the star in AU, $M_\star$ is the mass of the
star in $M_\odot$, and the parameter $\alpha$ is the
ratio of the radiation pressure force to stellar gravity
acting on the dust particles which is a strong function of
their size with smaller particles being more affected
(e.g., Wyatt et al. 1999).
Thus a disc that is 200-540 AU from the 2.3 solar mass star
HD99803B would have a P-R drag lifetime for the smallest
remnant grains ($\alpha = 1$) of 7-50 Myr.
As this is shorter than the age of the system of
$\sim 120$ Myr, for the disc we detected to be primordial
(i.e., not second generation), and so for its constituent dust
to have survived at this distance from the star over the age of
the system, this dust must have $\alpha \ll 0.1$ corresponding
to grains much larger than $\sim 0.1$ mm;
i.e., there can be no small ($<0.1$ mm) primordial grains
in this cold disc.\footnote{For the $1.6M_\odot$ star
HD112412 dust in its disc has a P-R drag lifetime of 5-12 Myr.
Similar conclusions to those for the HD99803B disc are
less certain due to the uncertainty in the age of this
system.}
On the other hand, such small grains are expected to be
abundant in a debris disc, since they are continually
replenished by the collisional destruction of larger
planetesimals that have longer P-R drag lifetimes.

Thus, while we are not proposing that the size distribution
of grains in primordial and debris discs should be inherently
different, we note that if these discs are protoplanetary,
their necessary lack of small ($<0.1$ mm) grains would have
a potentially discernable effect on their SEDs:
if protoplanetary the discs' emission would resemble black body
emission with $\beta = 0$ (e.g., Figure 3),
whereas if debris discs, their emission would have a $\beta$
closer to 1.
Thus it may be possible to to use a shallow sub-mm spectral
slope (i.e., emission $\propto 1/\lambda^{<2.5}$) to infer a
lack of small grains (Dent et al. 1998; Sheret et al. in
preparation) and a protoplanetary origin.
Further information about grain size could also be derived
from SED modelling if the radial location of the dust could
be determined by imaging the disc emission
(e.g., Wyatt \& Dent 2002).

\subsection{Evolution of Dust Mass}

\begin{figure}
  \begin{center}
    \begin{tabular}{rl}
      \textbf{(a)} & \hspace{-0.5in} \psfig{figure=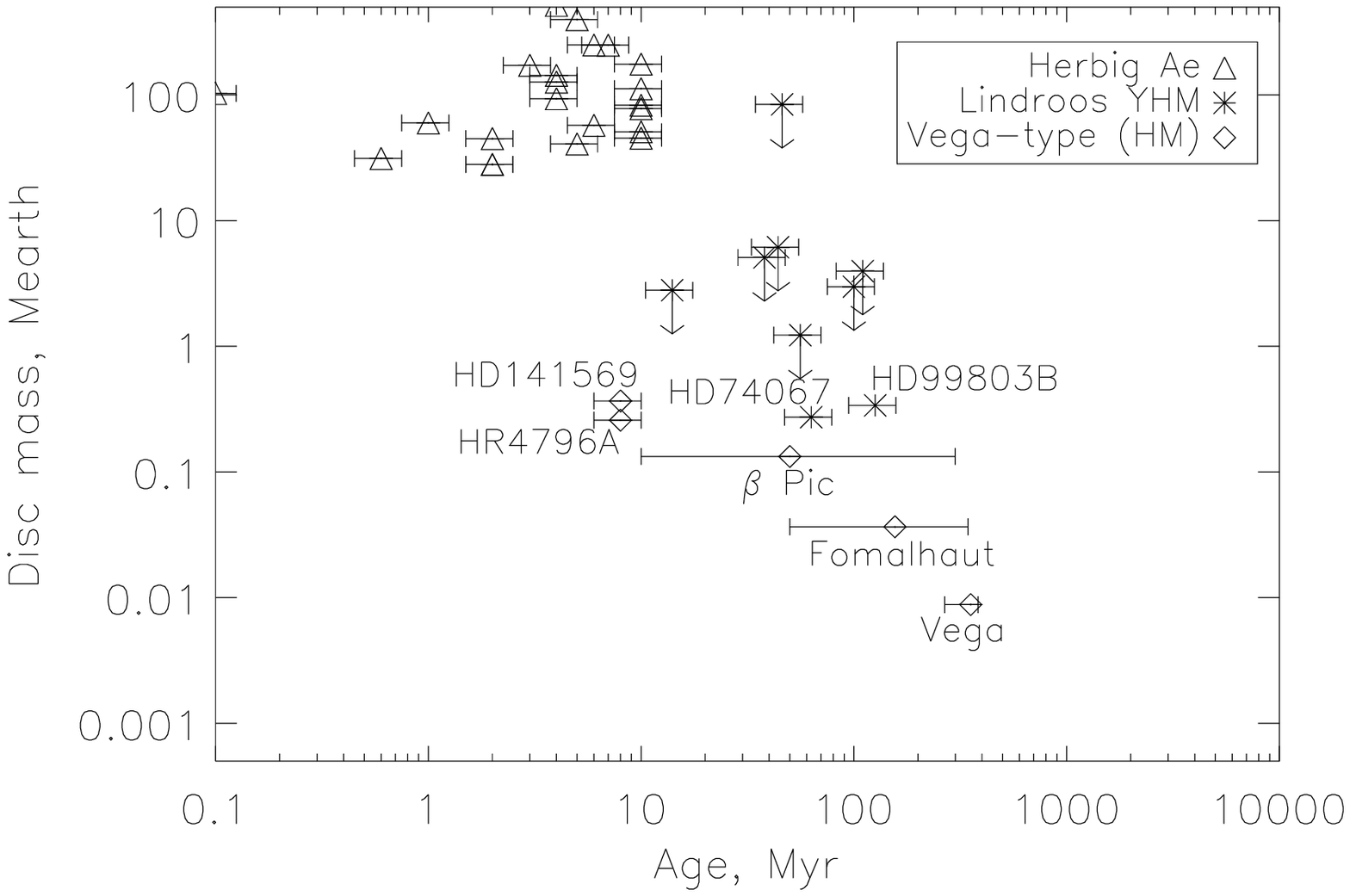,height=2.3in} \\
      \textbf{(b)} & \hspace{-0.5in} \psfig{figure=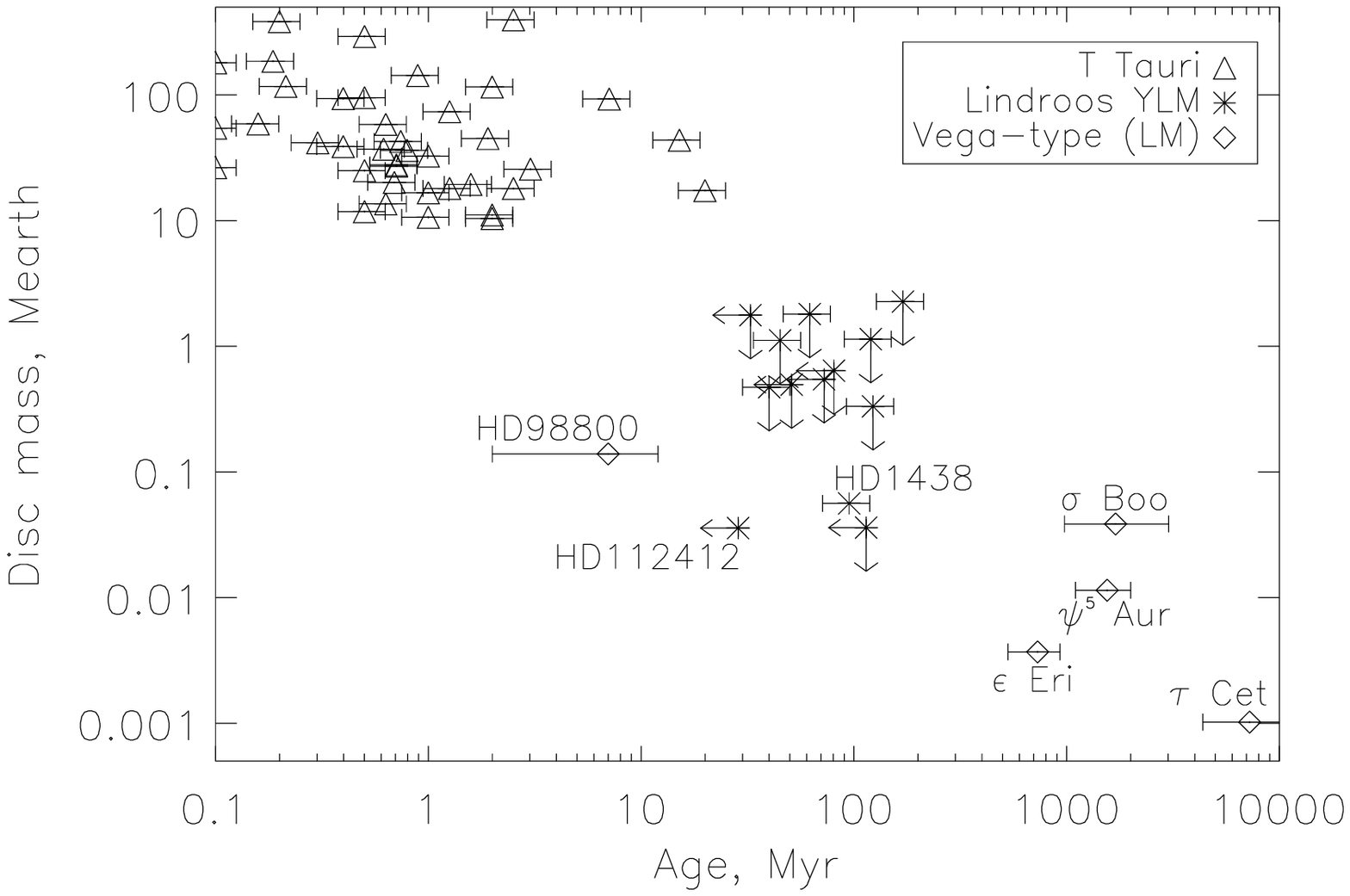,height=2.3in} \\
      \textbf{(c)} & \hspace{-0.5in} \psfig{figure=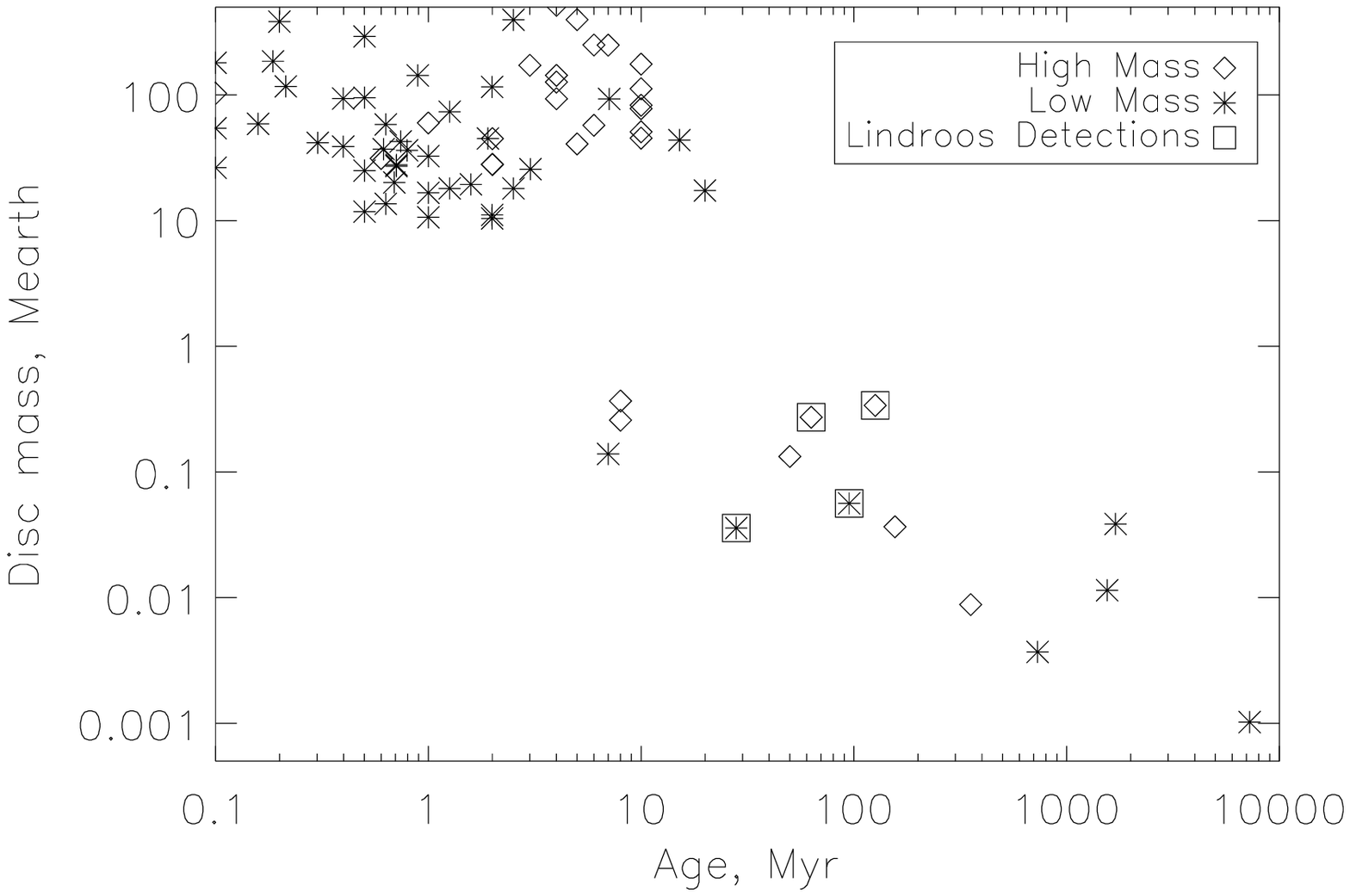,height=2.3in}
    \end{tabular}
    \caption{The evolution of dust mass for low mass \textbf{(a)} and
    high mass \textbf{(b)} stars.
    For all stars plotted, dust masses were calculated using equation (1) and
    published sub-mm or mm fluxes as well as estimates of the dust
    temperature and system distance.
    As well as the results of this study, we also show
    detections of T Tauri discs (Jewitt 1994; Osterloh \& Beckwith
    1995; N\"{u}rnberger, Chini \& Zinnecker 1997; N\"{u}rnberger et
    al. 1998) and of Herbig AeBe discs (Mannings \& Sargent 1997;
    and lists compiled by Natta et al. 1997 and Meeus et al. 2001);
    the errors in the ages are nominally plotted at $\pm 25$\%.
    Also plotted are the Vega-type stars for which sub-mm
    observations have confirmed the \textit{IRAS} excess to originate
    in a circumstellar disc (Holland et al. 1998; Greaves et al.
    1998; Sylvester et al. 2001; papers by Wyatt, Greaves, Sheret,
    et al. in preparation),
    and for which ages, with appropriate errors, are available in the
    literature (Lachaume et al. 1999; Song et al. 2000; Song et al. 2001).
    The bottom figure \textbf{(c)} combines the results shown in
    \textbf{(a)} and \textbf{(b)}, and for clarity omits the 
    the Lindroos non-detections and the error in the ages.}
    \label{fig:massevol}
  \end{center}
\end{figure}

Leaving aside the uncertainty in the origin of the emission we
detected, in Figure 4 we have plotted the dust masses
from Table 2 against system age assuming that the emission
is circumstellar.
This shows how the dust masses derived in this study compare
with those of the discs detected around stars of different ages.
While the dust mass limits achieved here vary
with system distance, for all but the most distant high
mass star in our sample we have been able to rule out the
presence of dust masses even an order of magnitude below
the levels that are characteristic of younger ($< 10$ Myr)
pre-main sequence (T Tauri or Herbig Ae) stars.
The dust masses that were detected in this sample, on the
other hand, are similar to those of the most massive debris
discs.
Thus there is no evidence that massive protoplanetary discs
persist far beyond $\sim 10$ Myr.

Figure 4c also emphasizes how protoplanetary and
debris discs appear to come from distinct populations,
since there are no known discs with masses in the
1-10$M_\oplus$ range.
This implies that the evolution from protoplanetary to
debris disc occurs rapidly at an age of $\sim 10$ Myr.
However, the apparent gap in the plot could in fact be
caused by a combination of the plot strategy and the
limited number of disc masses available to be plotted.
What Figure 4 really shows is the maximum disc
mass that we know exists around stars of different ages.
To interpret this correctly we also need to know the
fraction of stars at a given age that are plotted in this
figure.
Studies of young ($<10$ Myr) stars are biased toward either
classical or weak-line T Tauris (CTTs and WTTs respectively)
found in the nearest sites of recent star formation
(at $\sim 150$ pc).
The ratio of these populations (WTT/CTT) is uncertain, but
could be as high as 10 (Stahler \& Walter 1993).
The fraction of CTTs and WTTs that have $>10M_\oplus$ discs
is $\sim 50$\% and $\sim 10$\% respectively (Osterloh \&
Beckwith 1995), and the disc fraction for CTTs (but not that
of WTTs) may be higher, up to 100\%, when this mass limit is
decreased to $>2M_\oplus$ (Duvert et al. 2000).
Thus the total fraction of all young ($<10$ Myr) stars with massive
($>2M_\oplus$) discs could be as low as 18\%
(i.e., $[10\%*10+100\%*1]/11$).
Given that we would expect disc masses for collisionally
replenished discs to decline with age $\propto t^{-2}$ as
observed (Spangler et al. 2001), we thus might expect that,
if this decline starts at $\sim 10$ Myr, then by 70 Myr
(the average age of the Lindroos sample) 18\% of stars would
have discs more massive than $0.04 M_\oplus$.
This is consistent with our detection of $\geq 4/22$ discs
in the Lindroos sample at this level, and so from this study
alone we cannot state that the disc mass decline after 10 Myr
must be steeper than $t^{-2}$.

Rather it is possible that if CTTs and WTTs come from
separate populations, and do not form an evolutionary
sequence in which CTTs evolve into WTTs in short $\sim 0.1$ Myr
timescales (Skrutskie et al. 1990; Clarke et al. 2001),
then a large number of evolved CTTs could have
discs with dust masses in the range 1-10 $M_\oplus$ and these
would be detected once a larger sample of nearby young stars
has been observed.
Indeed we may already have observed one, since if the emission
toward HD74067 is bound to this system, then if this system
had been twice as distant (i.e., at a similar distance to the
nearest $<10$ Myr stars), the larger contribution of the
circumbinary material in central bolometer could have
resulted in the inference of a $1-10M_\oplus$ disc.
Apart from this Lindroos sample, suitable young ($\sim 10$ Myr)
nearby ($<70$ pc) candidates include members of the
$\beta$ Pictoris moving group (Zuckerman et al. 2001a)
and the TW Hydrae Association (Zuckerman et al. 2001b;
Zuckerman et al., in preparation).

\section{Conclusions}
Of the 22 young stars in our sample from the Lindroos catalog,
we report the detection of sub-mm emission (detected by us
using SCUBA) toward three of these stars and far-IR emission
(detected by \textit{IRAS}) toward another star:
\begin{itemize}

\item Two of the sub-mm detections (HD112412 and HD99803B)
we attribute to cold ($<40$ K assuming $\beta=1$) circumsecondary discs,
since these were not detected in the far-IR by \textit{IRAS} and
the primary falls outside the beam;
these discs have masses of $\sim 0.04$ and $0.3M_\oplus$,
respectively.
We were also able to show that a cold ($<50$ K assuming $\beta=0.5$)
circumprimary disc is present around HD112413 with a similar mass to the
circumsecondary HD112412 disc. 
These detections imply that current estimates of disc fractions
could be underestimated by a factor of two and indicate that
a significant cold submillimetre disc population could be awaiting
discovery in future sub-mm surveys.
It may be possible to determine whether these discs are
protoplanetary remnants or debris discs once their far-IR
emission and sub-mm spectral slopes have been measured,
since the emission, if protoplanetary,
would more closely resemble black body emission.

\item The sub-mm emission detected toward the third binary
system (HD74067) appears to be centrally peaked and extends out to
$\sim 70$ arcsec from the system.
The low temperature and high density of this emission means
that it is unlikely to be caused by local cirrus heated by the 
star.
Thus this is either a system containing a cold $0.3M_\oplus$
circumstellar disc as well as a $>27M_\oplus$ circumbinary
disc/envelope extending out to $\sim 6000$ AU, or this is a chance
alignment with an unrelated background object.
The former interpretation would support the idea that
T Tauri discs in binary systems are replenished by extended
circumbinary envelopes (Prato \& Simon 1997).
Since we were chopping onto an extended source, it was difficult
to determine the true structure of this emission, thus
further mapping of this region will help determine its
origin.
If this emission is bound to the star this would be an
extremely unusual system in that it retains a dust mass
of $\gg 1M_\oplus$ until an age of $\sim 60$ Myr.

\item The far-IR emission detected toward HD1438 implies that
one of the stars in this system harbours a 107 K, 0.05 $M_\oplus$
circumstellar disc.
If circumprimary, this disc could be resolved with mid-IR imaging
from an 8 metre telescope.

\end{itemize}

The low inferred dust masses for this sample supports the picture
that  protoplanetary dust discs are depleted to the levels of the
brightest debris discs ($\sim 1 M_\oplus$) at the end of 10 Myr.
However, a larger sample of young stars must be observed before
anything more concrete can be said on this subject.

\section*{Acknowledgments}
The JCMT is operated by the Joint Astronomy Centre,
on behalf of the UK Particle Physics and Astronomy
Research Council, the Netherlands Organization for
Pure Research, and the National Research Council
of Canada.
This research has made use of the NASA/ IPAC Infrared Science
Archive, which is operated by the Jet Propulsion Laboratory,
California Institute of Technology, under contract with the
National Aeronautics and Space Administration.
We would also like to thank an anonymous referee for
helpful comments.



\begin{thebibliography}{}
  \bibitem[\protect\citename{Andre \& Montmerle }1994]{am94}
     Andre P., Montmerle T., 1994, ApJ, 420, 837
  \bibitem[\protect\citename{Archibald et al.\ }2002]{ajhc02}
     Archibald E. N., et al., 2002, MNRAS, 336, 1
  \bibitem[\protect\citename{Artymowicz \& Lubow }1994]{al94}
     Artymowicz P., Lubow S. H., 1994, ApJ, 421, 651
  \bibitem[\protect\citename{Artymowicz \& Clampin }1997]{ac97}
     Artymowicz P., Clampin M., 1997, ApJ, 490, 863
  \bibitem[\protect\citename{Backman \& Paresce }1993]{bp93}
     Backman D. E., Paresce F., 1993, 
     in Levy E. H., Lunine J. I., eds., Protostars and Planets III.
     Univ. of Arizona Press, Tucson, p. 1253
  \bibitem[\protect\citename{Bate \& Bonnell }1997]{bb97}
     Bate M. R., Bonnell I. A., 1997, MNRAS, 285, 33
  \bibitem[\protect\citename{Beckwith et al. }1990]{bscg90}
     Beckwith S. V. W., Sargent A. I., Chini R. S., Guesten R.,
     1990, AJ, 99, 924
  \bibitem[\protect\citename{Bohren \& Huffman }1983]{bh83}
     Bohren C. F., Huffman D. R., 1983, Absorption and Scattering
     of Light by Small Particles. Wiley, New York
  \bibitem[\protect\citename{Clarke, Gendrin \& Sotomayor }2001]{cgs01}
     Clarke C. J., Gendrin A., Sotomayor M., 2001, MNRAS, 328, 485
  \bibitem[\protect\citename{Dent et al.\ }1998]{dmw98}
     Dent W. R. F., Matthews H. M., Ward-Thompson D., 1998,
     MNRAS, 301, 1049
  \bibitem[\protect\citename{Dent et al.\ }2000]{dwhg00}
     Dent W. R. F., Walker H. J., Holland W. S., Greaves J. S., 2000,
     MNRAS, 314, 702
  \bibitem[\protect\citename{Duquennoy \& Mayor }1991]{dm91}
     Duquennoy A., Mayor M., 1991, A\&A, 248, 485
  \bibitem[\protect\citename{Duvert et al.\ }2000]{duv00}
     Duvert G., Guilloteau S., M\'{e}nard F., Simon M., Dutrey A.,
     2000, A\&A, 355, 165
  \bibitem[\protect\citename{Eales et al.\ }2000]{eal00}
     Eales S., Lilly S., Webb T., Dunne L., Gear W., Clements D.,
     Yun M., 2000, AJ, 120, 2244
  \bibitem[\protect\citename{Gahm et al.\ }1983]{gal83}
     Gahm G. F., Ahlin P., Lindroos K. P., 1983, A\&AS, 51, 143
  \bibitem[\protect\citename{Gahm et al.\ }1994]{gzpp94}
     Gahm G. F., Zinnecker H., Pallavicini R., Pasquini L., 1994,
     A\&A, 282, 123
  \bibitem[\protect\citename{Gaustad \& van Buren }1993]{gv93}
     Gaustad J. E., Van Buren D., 1993, PASP, 105, 1127
  \bibitem[\protect\citename{Gerbaldi et al.\ }2001]{gfb01}
     Gerbaldi M., Faraggiana R., Balin N., 2001, A\&A, 379, 162
  \bibitem[\protect\citename{Greaves et al.\ }1998]{gh98}
     Greaves J. S., et al., 1998, ApJ, 506, L133
  \bibitem[\protect\citename{Greaves et al.\ }2000]{gmh00}
     Greaves J. S., Mannings V., Holland W. S., 2000, Icarus, 143, 155
  \bibitem[\protect\citename{Habing et al.\ }1999]{hdjk99}
     Habing H. J., et al., 1999, Nature, 401, 456
  \bibitem[\protect\citename{Haisch, Lada \& Lada }2001]{hll01}
     Haisch K. E., Lada E. A., Lada C. J., 2001, ApJ, 553, L153
  \bibitem[\protect\citename{Herbig }1978]{herb78}
     Herbig G. H., 1978, in Mirzoya L. V., ed., Problems of Physics
     and Evolution of the Universe.
     Armenian Acad. Sci., Yerevan, p. 171
  \bibitem[\protect\citename{Holland et al.\ }1998]{hgzw98}
     Holland W. S., et al., 1998, Nature, 392, 788
  \bibitem[\protect\citename{Holland et al.\ }1999]{holl99}
     Holland W. S., et al., 1999, MNRAS, 303, 659
  \bibitem[\protect\citename{Hollenbach et al. }2000]{holl00}
     Hollenbach D., Yorke H. W., Johnstone D., 2000,
     in Mannings V., Boss A. P., Russell S. S., eds, Protostars
     \& Planets IV.
     Univ. Arizona Press, Tucson, p. 401
  \bibitem[\protect\citename{Huelamo et al.\ }2001]{hbbn01}
     Hu\'{e}lamo N., Brandner W., Brown A. G. A., Neuh\"{a}user R.,
     Zinnecker H., 2001, A\&A, 373, 657
  \bibitem[\protect\citename{Huelamo et al.\ }2000]{hnss00}
     Hu\'{e}lamo N., Neuh\"{a}user R., Stelzer B., Supper R.,
     Zinnecker H., 2000, A\&A, 359, 227
  \bibitem[\protect\citename{Jenness \& Lightfoot }1998]{jl98}
     Jenness T., Lightfoot J. F., 1998, in Albrecht R., Hook R. N.,
     Bushouse H. A., eds., ASP Conf. Ser. Vol. 145, Astronomical Analysis
     Software and Systems VII. Astron. Soc. Pac., San Francisco, p. 216
  \bibitem[\protect\citename{Jensen \& Akeson }2003]{ja03}
     Jensen E. L. N., Akeson R. L., 2003, ApJ, 584, 875
  \bibitem[\protect\citename{Jensen, Mathieu \& Fuller }1996]{jmf96}
     Jensen E. L. N., Mathieu R. D., Fuller G. A., 1996, ApJ, 458, 312
  \bibitem[\protect\citename{Jewitt }1994]{jewi94}
     Jewitt D. C., 1994, AJ, 108, 661
  \bibitem[\protect\citename{Johnstone, Hollenbach \& Bally }1998]{jhb98}
     Johnstone D., Hollenbach D., Bally J., 1998, 499, 758
  \bibitem[\protect\citename{Kalas et al.\ }2002]{kal02}
     Kalas P., Graham J. R., Beckwith S. V. W., Jewitt D. C.,
     Lloyd J. P., 2002, ApJ, 567, 999
  \bibitem[\protect\citename{Kenyon \& Hartmann }1995]{kh95}
     Kenyon S. J., Hartmann L., 1995, ApJS, 101, 117
  \bibitem[\protect\citename{Lachaume et al. }1999]{ldlh99}
     Lachaume R., Dominik C., Lanz T., Habing H. J., 1999, A\&A, 348, 897
  \bibitem[\protect\citename{Lagrange et al. }2000]{lagr00}
     Lagrange A.-M., Backman D. E., Artymowicz P., 2000,
     in Mannings V., Boss A. P., Russell S. S., eds, Protostars
     \& Planets IV.
     Univ. Arizona Press, Tucson, p. 639
  \bibitem[\protect\citename{Lindroos }1985]{lind85}
     Lindroos K. P., 1985, A\&AS, 60, 183
  \bibitem[\protect\citename{Lindroos }1986]{lind86}
     Lindroos K. P., 1986, A\&A, 156, 223
  \bibitem[\protect\citename{Lissauer }1993]{liss93}
     Lissauer J. J., 1993, ARA\&A, 31, 129
  \bibitem[\protect\citename{Mannings \& Sargent }1997]{ms97}
     Mannings V., Sargent A. I., 1997, ApJ, 490, 792
  \bibitem[\protect\citename{Martin }1997]{mart97}
     Mart\'{i}n E. L., 1997, A\&A, 321, 492
  \bibitem[\protect\citename{Martin et al.\ }1992]{mmr92}
     Mart\'{i}n E. L., Magazz\`{u} A., Rebolo R., 1992, A\&A, 257, 186
  \bibitem[\protect\citename{Matsuyama et al.\ }2002]{mjh02}
     Matsuyama I., Johnstone D., Hartmann L., 2003, ApJ, 582, 893
  \bibitem[\protect\citename{Meeus et al.\ }2001]{mwbv01}
     Meeus G., Waters L. B. F. M., Bouwman J., van den Ancker M. E.,
     Waelkens C., Malfait K., 2001, A\&A, 365, 476
  \bibitem[\protect\citename{Murphy }1969]{murp69}
     Murphy R. E., 1969, AJ, 74, 1082
  \bibitem[\protect\citename{Natta et al. }1997]{ngmu97}
     Natta A., Grinin V. P., Mannings V., Ungerechts H., 1997, ApJ, 491, 885
  \bibitem[\protect\citename{N\"{u}rnberger et al. }1997]{ncz97}
     N\"{u}rnberger D., Chini R., Zinnecker H., 1997, A\&A, 324, 1036
  \bibitem[\protect\citename{N\"{u}rnberger et al. }1998]{ncz98}
     N\"{u}rnberger D., Brandner W., Yorke H. W., Zinnecker H., 1998,
     A\&A, 330, 549
  \bibitem[\protect\citename{O'Dell }2001]{odel01}
     O'Dell C. R., 2001, ARA\&A, 39, 99
  \bibitem[\protect\citename{Osterloh \& Beckwith }1995]{ob95}
     Osterloh M., Beckwith S. V. W., 1995, ApJ, 439, 288
  \bibitem[\protect\citename{Pallavicini et al.\ }1992]{ppr92}
     Pallavicini R., Pasquini L., Randich S., 1992, A\&A, 261, 245
  \bibitem[\protect\citename{Papaloizou \& Pringle }1977]{pp77}
     Papaloizou J., Pringle J. E., 1977, MNRAS, 181, 441
  \bibitem[\protect\citename{Pollack et al.\ }1994]{phbs94}
     Pollack J. B., Hollenbach D., Beckwith S., Simonelli D. P.,
     Roush T., Fong W., 1994, ApJ, 421, 615
  \bibitem[\protect\citename{Prato \& Simon }1997]{ps97}
     Prato L., Simon M., 1997, ApJ, 474, 455
  \bibitem[\protect\citename{Ray et al.\ }1995]{rsbk95}
     Ray T. P., Sargent A. I., Beckwith S. V. W., Koresko C., Kelly P.,
     1995, 440, L89
  \bibitem[\protect\citename{Scott et al.\ }2002]{scot02}
     Scott S. E. et al., 2002, MNRAS, 331, 817
  \bibitem[\protect\citename{Shu et al.\ }1987]{sal87}
     Shu, F. H., Adams, F. C., \& Lizano, S. 1987, ARA\&A, 25, 23
  \bibitem[\protect\citename{Skinner, Brown \& Walter }1991]{sbw91}
     Skinner S. L., Brown A., Walter F. M., 1991, AJ, 102, 1742
  \bibitem[\protect\citename{Skrutskie et al.\ }1990]{sdse90}
     Skrutskie M. F., Dutkevitch D., Strom S. E., Edwards S.,
     Strom K. M., Shure M. A., 1990, AJ, 99, 1187
  \bibitem[\protect\citename{Song et al.\ }2000]{scbs00}
     Song I., Caillault J.-P., Barrado Y Navascu\'{e}s D.,
     Stauffer J. R., Randich S., 2000, ApJ, 532, L41
  \bibitem[\protect\citename{Song et al.\ }2001]{scbs01}
     Song I., Caillault J.-P., Barrado Y Navascu\'{e}s D.,
     Stauffer J. R., 2001, ApJ, 546, 352
  \bibitem[\protect\citename{Spangler et al.\ }2001]{sssb01}
     Spangler C., Sargent A. I., Silverstone M. D., Becklin E. E.,
     Zuckerman B., 2001, ApJ, 555, 932
  \bibitem[\protect\citename{Stahler \& Walter }1993]{sw93}
     Stahler S. W., Walter F. M., 1993, 
     in Levy E. H., Lunine J. I., eds., Protostars and Planets III.
     Univ. of Arizona Press, Tucson, p. 405
  \bibitem[\protect\citename{Strom et al.\ }1989]{ssec89}
     Strom K. M., Strom S. E., Edwards S., Cabrit S., Skrutskie M. F.,
     1989, AJ, 97, 1451
  \bibitem[\protect\citename{Sylvester et al.\ }2001]{sdb01}
     Sylvester R. J., Dunkin S. K., Barlow M. J., 2001, MNRAS, 327, 133
  \bibitem[\protect\citename{Takeuchi \& Artymowicz }2001]{ta01}
     Takeuchi T., Artymowicz P., 2001, ApJ, 557, 990
  \bibitem[\protect\citename{Waters \& Waelkens }1998]{ww98}
     Waters L. B. F. M., Waelkens C., 1998, ARA\&A, 36, 233
  \bibitem[\protect\citename{Webb et al. }1999]{webb99}
     Webb R. A., Zuckerman B., Platais I., Patience J., White R. J.,
     Schwartz M. J., McCarthy C., 1999, ApJ, 512, L63
  \bibitem[\protect\citename{Weidenschilling \& Cuzzi }1993]{weid93}
     Weidenschilling S. J., Cuzzi J. N. 1993,
     in Levy E. H., Lunine J. I., eds, Protostars \& Planets III.
     Univ. Arizona Press, Tucson, p. 1031
  \bibitem[\protect\citename{White \& Ghez }2001]{wg01}
     White R. J., Ghez A. M., 2001, ApJ, 556, 265
  \bibitem[\protect\citename{Worley \& Douglass }1996]{wd96}
     Worley C. E., Douglass G. G., 1996, A\&AS, 125, 523
  \bibitem[\protect\citename{Wyatt \& Dent }2002]{wd02}
     Wyatt M. C., Dent W. R. F., 2002, MNRAS, 334, 589     
  \bibitem[\protect\citename{Wyatt et al. }1999]{wdtf99}
     Wyatt M. C., Dermott S. F., Telesco C. M.,
     Fisher R. S., Grogan K., Holmes E. K., Pi\~{n}a R. K.,
     1999, ApJ, 527, 918
  \bibitem[\protect\citename{Zuckerman }2001]{zuck01}
     Zuckerman B., 2001, ARA\&A, 39, 549
  \bibitem[\protect\citename{Zuckerman \& Becklin }1993]{zb93}
     Zuckerman B., Becklin E. E., 1993, ApJ, 414, 793
  \bibitem[\protect\citename{Zuckerman et al. }2001]{zsbw01}
     Zuckerman B., Song I., Bessell M. S., Webb R. A., 2001a,
     ApJ, 562, L87
  \bibitem[\protect\citename{Zuckerman et al. }2001]{zsbw01}
     Zuckerman B., Webb R. A., Schwart M., Becklin E. E., 2001b,
     ApJ, 549, L233
\end{thebibliography}
\end{document}